\documentclass[aip,graphicx,amsmath, amssymb, reprint]{revtex4-1}

\usepackage{graphicx}
\usepackage{dcolumn}
\usepackage{bm}
\usepackage[dvipsnames]{xcolor}

\usepackage[utf8]{inputenc}
\usepackage[T1]{fontenc}
\usepackage{mathptmx}
\usepackage{etoolbox}

\makeatletter
\def\@email#1#2{%
	\endgroup
	\patchcmd{\titleblock@produce}
	{\frontmatter@RRAPformat}
	{\frontmatter@RRAPformat{\produce@RRAP{*#1\href{mailto:#2}{#2}}}\frontmatter@RRAPformat}
	{}{}
}%
\makeatother

\draft 

\graphicspath{{./Images/}}
\draft 

\DeclareMathOperator*{\E}{\mathbb{E}}
	
\begin{document}
\title[]{Discrete state model of a self-aggregating colloidal system with directional interactions} 
\author{Salman Fariz Navas}
\email[]{salman.fariz.navas@tu-berlin.de}
\author{Sabine H. L. Klapp}
\email[]{sabine.klapp@tu-berlin.de}
\affiliation{Institute for Theoretical Physics, Technical University of Berlin, Hardenbergstr. 36, 10623 Berlin, Germany}
\date{\today}
		
\begin{abstract}
The construction of coarse-grained descriptions of a system's kinetics is well established in biophysics. One prominent example is Markov state models in protein folding dynamics. In this paper, we develop a coarse-grained, discrete state model of a self-aggregating colloidal particle system inspired by the concepts of Markov state modelling. The specific self-aggregating system studied here involves field-responsive colloidal particles in orthogonal electric and magnetic fields. Starting from particle-resolved (Brownian dynamics) simulations, we define the discrete states by categorizing each particle according to it's local structure. We then describe the kinetics between these states as a series of stochastic, memoryless jumps. In contrast to other works on colloidal self-assembly, our coarse-grained approach describes the simultaneous formation and evolution of multiple aggregates from single particles. Our discrete model also takes into account the changes in transition dynamics between the discrete states as the size of the \textit{largest} cluster grows. We validate the coarse-grained model by comparing the predicted population fraction in each of the discrete states with those calculated directly from the particle-resolved simulations as a function of the largest cluster size. We then predict population fractions in presence of noise-averaging and in a situation where a model parameter is changed instantaneously after a certain time. Finally, we explore the validity of the detailed balance condition in the various stages of aggregation. 
\end{abstract}
		
\maketitle 

\section{Introduction} 
Self-aggregating colloidal systems \cite{hong2006clusters, hong2008clusters, klokkenburg2006quantitative, camp2000isotropic, sacanna2010lock,chen2011directed, duguet2011design, yang2013tunable, wang2012colloids, groschel2013guided} display a rich variety of aggregated structures depending on the character of the underlying inter-particle interactions. In many cases, these interactions are direction-dependent. Examples of the resulting manifold of structures \cite{mao2013entropy, van2014entropically,hong2006clusters, hong2008clusters, sciortino2011reversible, russo2009reversible, de2006dynamics} are chains \cite{sacanna2010lock,oh2019colloidal,bharti2015assembly, bharti2016multidirectional,mallory2018active}, rings \cite{liao2020dynamical, oh2019colloidal}, system spanning networks \cite{sciortino2011reversible, russo2009reversible,mallory2019activity, wei2023reconfiguration, oh2019colloidal} and also irregularly shaped clusters \cite{kogler2015generic, lander2013crystallization, grober2023unconventional}. Direction-dependent interactions can arise, for example, by decorating the particle's surface with hydrophobic patches \cite{chen2011directed, hong2008clusters, hong2006clusters, mallory2017self, glotzer2007anisotropy} or via dipolar interactions \cite{teixeira2000effect, klokkenburg2006quantitative, camp2000isotropic, bharti2015assembly, bharti2016multidirectional, kogler2015generic}. During the process of the aggregation, the particles typically first form multiple smaller clusters or "molecules" which then merge into larger structures. Controlling the sizes, shapes and internal structures of these aggregates is a topic that attracts researchers from various fields, ranging from physics and chemistry \cite{yin2001template} to material science \cite{song2019large} and biology \cite{vernerey2019biological,mueller2019importance}. In recent years, there is increasing interest to understand not only the various final (steady-state) structures that the particles aggregate into, but also the dynamical processes during the course of aggregation, even when the final state corresponds to thermal equilibrium. Here, computer simulations of suitable model systems seem to be the method of choice. However, in particle-resolved simulations, the timescales required for the formation of final structures are typically orders of magnitudes larger than the timescales accessible computationally. Therefore, a promising approach is to develop coarse-grained models for self-aggregation that focus on structural transitions between aggregates. In recent years, different coarse-graining techniques have been used to study colloidal self-aggregation. Some notable examples include the string method \cite{pan2008finding, weinan2002string, ovchinnikov2011free}, metadynamics \cite{laio2002escaping, iannuzzi2003efficient} and cluster-move Monte Carlo techniques \cite{whitelam2009role}. A major limitation of these methods; however, is that they typically require an \textit{a priori} knowledge of the energy landscape and are limited to resolving only a few of the many different aggregation pathways \cite{perkett2014using}. An alternative route is inspired from Markov state models (MSM) for protein folding dynamics \cite{bowman2013introduction,prinz2011markov,husic2018markov}. Indeed, recent studies have demonstrated the applicability of MSMs also for self-aggregating colloidal systems \cite{appeldorn2021employing, perkett2014using, yang2022nanoparticle,lander2013crystallization,trubiano2022optimization,tang2017construction,sengupta2019automated, trubiano2024markov}.  

Many of these approaches follow a similar workflow in the sense that the discrete Markov states entering the MSM are constructed from particle-resolved simulations, such as Molecular Dynamics, of the underlying system. In most applications of MSM, be it protein folding or colloidal self-aggregation, one is interested in the dynamical evolution of a single molecule (or cluster). This allows to perform the discretization of the MSM on the global configuration space \cite{appeldorn2021employing,husic2018markov,bowman2013introduction,perkett2014using, yang2022nanoparticle, trubiano2022optimization, tang2017construction, sengupta2019automated}; that is, each discrete state involves a configuration of the entire system at a given instant of time. This approach was used, for example, to study the assembly of a small isolated colloidal cluster \cite{trubiano2022optimization, tang2017construction, sengupta2019automated}. However, such an approach is not directly applicable for larger systems with hundreds (or even thousands) of particles which can self-aggregate simultaneously into multiple clusters \cite{kogler2015generic, bharti2015assembly, bharti2016multidirectional, grober2023unconventional, mallory2017self, mallory2018active, mallory2019activity} and then merge or dissolve. Here, the transition rates will also be time-dependent since the number of "free" non-aggregated particles changes with time. Recent studies have introduced the concept of "MultiMSMs" to study such systems \cite{trubiano2024markov}. In this context, discretization is performed on the local configurational space of a particular aggregate within a large system. Each aggregate is then treated as a quasi-independent local subsystem (referred to as the independent Markov decomposition (IMD) method) \cite{hempel2021independent, trubiano2024markov}.  

In the present paper, we develop a different discrete state model to study the self-aggregation of a two-dimensional (2D) colloidal system. The underlying model describes aggregation of polarizable particles in orthogonal electric and magnetic fields \cite{kogler2015generic, navas2024impact}. In contrast to the MultiMSM approach described in Ref. \cite{trubiano2024markov}, where the system involves the aggregation of larger subunits consisting of rigid bodies (pentagonal and triangular) into dodecahedral capsids, the subunits in our self-aggregating system consists of single Brownian colloidal particles. The underlying interaction potential is anisotropic, with the degree of anisotropy being a tunable parameter. As shown in an earlier study \cite{kogler2015generic} based on Brownian dynamics (BD) simulations, this model allows for the formation of aggregates with different structures. Specifically, one observes hexagonal ordered clusters at low anisotropy, square-like clusters at high anisotropy and a co-existence of both types of structures at intermediate anisotropy \cite{kogler2015generic}. Our primary focus is to coarse-grain the dynamics of local structural transitions as aggregation proceeds. Hence, in the present study, we define the discrete states by identifying the prominent "local" aggregate structures observed in the BD simulations and categorize them using suitable order parameters. Inspired by a recent study of crystallization of Brownian particles under shear\cite{lander2013crystallization}, we handle the time-dependent nature of the transition rates via the largest cluster size, based on a time-scale separation between the state transitions and cluster growth times. We work under the IMD assumption (similar to, e.g., Ref.~\cite{trubiano2024markov, hempel2021independent}) and use transitions occurring in all of the different aggregates to construct the MSM. We validate our discrete state model by comparing it with particle-resolved BD simulation data. Further, we demonstrate the applicability of our method to simulate new "unseen" scenarios, where the parameter tuning the anisotropy is changed during the simulation. Finally, we identify uni-directional transitions in the early stages of aggregation, leading to a transient violation of the detailed balance condition on the coarse-grained scale.

The rest of the paper is organised as follows. In Section~\ref{section: particle_based}, we begin with the particle-resolved simulations of the self-aggregating colloidal system, describing first, the model and then summarizing key results from Ref. \cite{kogler2015generic}. We also define the order parameters that are used to identify the different structures. In Section~\ref{section: coarse_grained}, we describe step-by-step the construction of the coarse-grained discrete state model. In Section~\ref{results}, we show predictions of the discrete state model, including a validation of the accuracy by comparison with particle-resolved simulation data and an investigation of aggregation after parameter switching. Further, we investigate deviations from the detailed balance condition and their connection to uni-directional transitions on the coarse-grained level. Finally, we summarize our findings in Section~\ref{section: conclusion}.

\section{Particle-based simulations} \label{section: particle_based}
\subsection{Model}  \label{model}
We study a two dimensional (2D) colloidal system consisting of $N$ disk-like particles with diameter $\sigma$. The position of each particle $i$ ($i=1,....,N$) evolves according to the Brownian dynamics (BD) equation, 
\begin{equation}
	\gamma\dot{\mathbf{r}}_i=-\sum_{j\neq i}\nabla_{\mathbf{r}_{i}} U(\mathbf{r}_{ij})+\mathbf{F}_{R,i}(t). \label{eq:BD}
\end{equation} 
	
In Eq.~(\ref{eq:BD}) $\mathbf{F}_{R,i}(t)$ is a random force acting on particle $i$ at time $t$ with the properties of a Gaussian white noise, that is, $\left< F^a_{R,i}(t) \right>=0$ and $\left<F^a_{R,i}(t)  F^b_{R,j}(t') \right>=2\gamma k_BT \delta(t-t')\delta_{ij}\delta_{ab}$ where $i, j$ are particle indices and $a,b \in \left\{ 1,2\right\}$ represent the vector components. Further, $\gamma$ is the friction coefficient of the medium in which the colloidal particles are dispersed in, $T$ is the temperature and $k_B$ the Boltzmann constant. The vector $\mathbf{r}_{ij}$ is the centre-to-centre displacement vector between particles $i$ and $j$. The interaction potential $U(\mathbf{r}_{ij})$ has two contributions,
	
\begin{equation}
	U(\mathbf{r}_{ij})=U_{\text{LJ}}(r_{ij})+U_{\text{DIP}} (\mathbf{r}_{ij}) \label{eq:full_U}.
\end{equation}

\begin{figure*} [htp]
	\centering
	\includegraphics[width=1.0\textwidth]{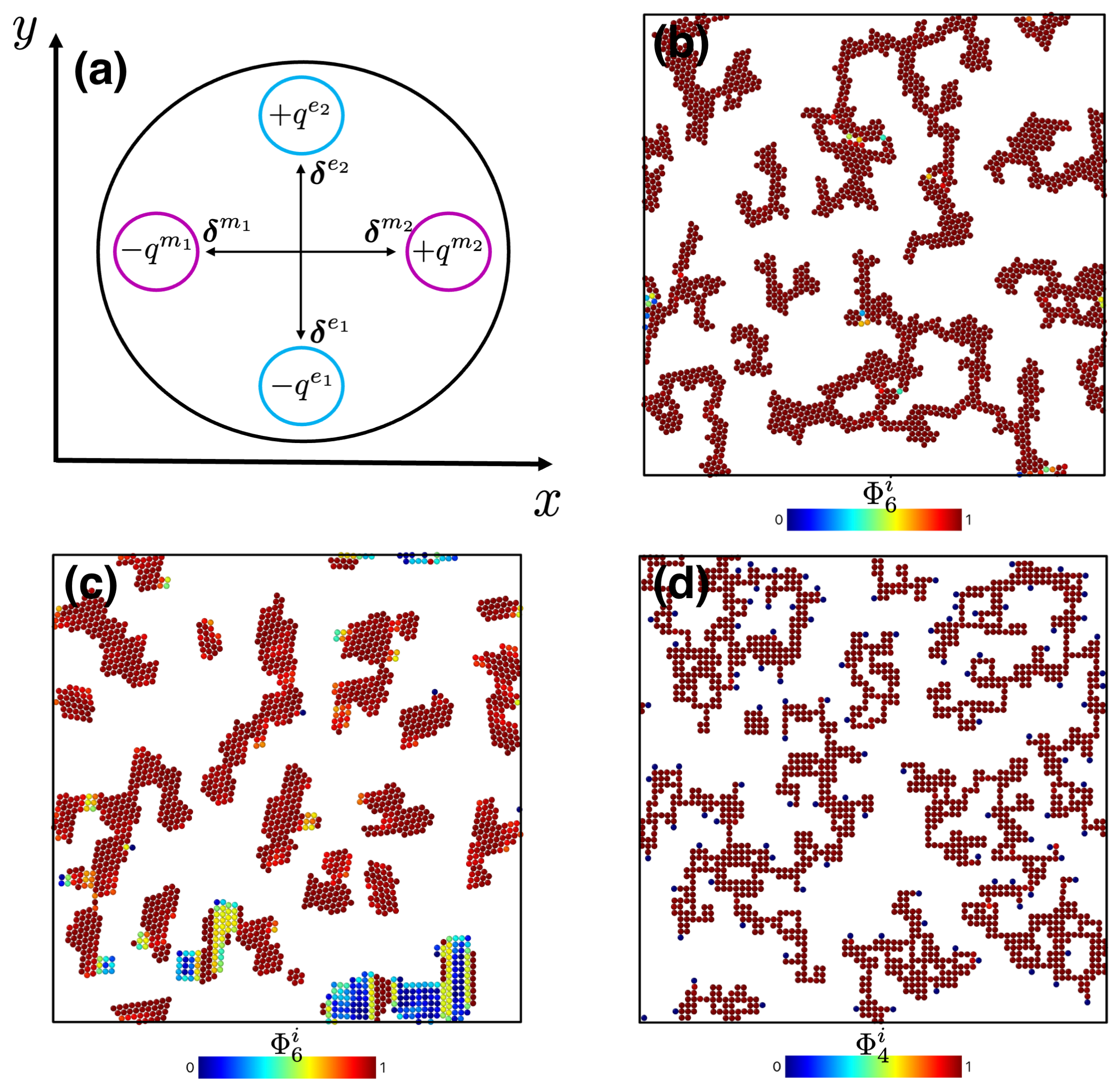}
	\caption{(a) Schematics of a particle with two pairs of fictitious point charges mimicking induced dipole moments. The point charges are colour coded according to the corresponding dipole type: blue for electric field ($e$) pointing in $y-$direction and purple for the magnetic field ($m$) pointing in $x-$direction. The position of the charges are shifted away from the particle centre by a vector $\boldsymbol{\delta}^{\alpha_k} \in \left[ -\delta_r \mathbf{\hat{x}},\delta_r\mathbf{\hat{x}}, -\delta_r \mathbf{\hat{y}}, \delta_r \mathbf{\hat{y}}  \right]$ with $\alpha=\text{e, m}$ and $k=1, 2$ pointing either parallel or anti-parallel to the corresponding field. Simulation snapshots (after $100\tau_B$) at three values of the charge separation parameter: (b) case~1: $\delta_r=0.1\sigma$, (c) case~2: $\delta_r=0.21\sigma$, (d) case~3: $\delta_r=0.3\sigma$. The other parameters are set to $\rho^{*}=0.3$ and $T^{*}=0.05$.} \label{fig:BD_model}
\end{figure*}
	
The first term on the right side of Eq.~(\ref{eq:full_U}) represents a shifted and truncated Lennard-Jones potential (commonly referred to as the Weeks-Chandler-Andersen (WCA) potential) for all particle pairs $i\neq j$ \cite{wca}. In order to make the potential $U_{\text{LJ}}(r_{ij})$ repulsive everywhere, we employ a cut-off distance of $r_{\text{LJ}}=2^{1/6}\sigma$. 
	
The second term on the right side of Eq.~(\ref{eq:full_U}) is the direction-dependent potential $U_{\text{DIP}} (\mathbf{r}_{ij})$. Our ansatz for this potential is inspired by earlier work on modelling field-responsive particles in orthogonal electric ($e$) and magnetic ($m$) fields \cite{kogler2015generic}. In a more recent study \cite{navas2024impact}, we had extended the same model to study the combined effect of non-reciprocal and anisotropic interactions.

Specifically, we assume that these external fields point along the $x-$ and $y-$axes and induce dipole moments $\boldsymbol{\mu}^{\alpha}$ with $\alpha=e, m$ along their respective directions. Each of these dipole moments is mimicked by two fictitious, opposite point charges $q^{\alpha_1}=- q^{\alpha_2}$ placed inside each particle. Hence, each particle comprises of two pairs of "charges": one pair representing the induced electric dipole moment and the other one representing the magnetic dipole moment as shown in Fig.~\ref{fig:BD_model}(a). The position of the charges relative to the particle centre is given by a vector $\boldsymbol{\delta}^{\alpha_k}=(-1)^k \delta_r \mathbf{\hat{e}}_{\alpha}$, with $k=1, 2$. Thus, the vectors related to the positive charges point parallel to the direction of the corresponding field ($\mathbf{\hat{e}}_\text{m}=\mathbf{\hat{x}}$ for magnetic field and $\mathbf{\hat{e}}_\text{e}=\mathbf{\hat{y}}$ for electric field) while those related to negative charges point anti-parallel. The parameter $\delta_r$ gives the distance between each of these point charges from the particle's centre and may be considered as a measure of the strength of the fields\cite{kogler2015generic}. Following Ref. \cite{kogler2015generic, navas2024impact}, we assume that each point charge has the same absolute value $|\pm q|=2.5\left(\epsilon/\sigma\right)^{1/2}$. The interaction between two point charges of different particles $i, j$ of the same type $\alpha=e, m$ and sign $k,l=1, 2$ is described by a Yukawa potential \cite{kogler2015generic},
	
\begin{equation}
	U^{\alpha_{k l}}_{\text{YU}}({r_{ij}^{\alpha_{kl}})}=q^{\alpha_k} q^{\alpha_l} \frac{\exp\left(-\kappa r_{ij}^{\alpha_{kl}}\right)}{r_{ij}^{\alpha_{k l}}}, \label{eq:yukawa}
\end{equation}
	
where $r_{ij}^{\alpha_{k l}}=|\mathbf{r}_{ij}+\boldsymbol{\delta}^{\alpha_l}-\boldsymbol{\delta}^{\alpha_k}|$  is the distance between the charge pairs. By using the Yukawa potential instead of a true Coulomb potential, we implicitly assume that all interactions are screened with an inverse screening length $\kappa$. Here we set $\kappa=4.0/\sigma$. The potential is cut-off at a distance $r_{\text{YU}}=4.0\sigma$. The overall contribution of the charges to the pair-interaction potential is given by the sum of the Yukawa potentials for the magnetic and electric charges. This yields $U_{\text{DIP}}(\mathbf{r}_{ij})=\sum_{\alpha\in \text{e,m}}\sum_{k,l=1}^{2} U^{\alpha_{kl}}_{\text{YU}}({r_{ij}^{\alpha_{kl}})}$. Note that we have normalised  $U_{\text{DIP}}(\mathbf{r}_{ij})$ such that it has a constant value for $r_{ij}=\sigma$ and $\mathbf{r}_{ij}$ pointing along one of the fields ($\mathbf{r}_{ij}=\sigma\hat{\mathbf{e}}_{\alpha}$) (see Appendix \ref{app_norm_udip}) for different values of $\delta_r$. This is done to facilitate the comparison between different values of $\delta_r$. 
	
To investigate the system's aggregation behaviour we perform simulations for $N=1800$ particles at different reduced number densities $\rho^*=\rho\sigma^2$ and temperatures $T^*=k_BT/\epsilon$ in a square-like simulation box with periodic boundary conditions. The BD equations of motion [see Eq.~(\ref{eq:BD})] are solved using the Euler-Maruyama integration scheme for stochastic ordinary differential equations\cite{kloeden1992stochastic}. The integration step-size is set to $\Delta t=10^{-5}\tau_B$, where $\tau_B$ is defined as the Brownian timescale of the system and is given as $\tau_B= \sigma^2\gamma/k_BT$. The simulations are performed until the simulation time reaches $100\tau_B$.

\subsection{Aggregation on the particle level} \label{aggregates_studied}
The present study builds on results from extensive BD simulations reported in Ref.~\cite{kogler2015generic, navas2024impact}, where the aggregation behaviour has been investigated for a range of densities, temperatures, and values of the charge separation parameter $\delta_r$. We here summarize some main findings. At the strengths of the "charges" considered here and densities in the range $\rho^{*}\lesssim 0.45$, the system assumes a disordered, fluid-like state at temperatures $T^{*}\gtrsim 0.4-0.6$ (depending on the precise value of $\delta_r$). At lower temperatures and densities in the range $\rho^{*}\lesssim 0.45$, the particles aggregate into isolated clusters whose structural characteristics depend on the charge distribution (see below). Further, at low temperatures and larger densities in the range $\rho^{*}=0.45-0.7$ one observes highly percolated structures where clusters of different types interconnect with each other forming a frustrated network.

In the present study we focus on the formation of isolated clusters in the range at density $\rho^{*}=0.3$ (yielding isolated clusters) and temperature $T^{*}=0.05$. Following ~\cite{kogler2015generic}, we distinguish different cluster types by the mean (i.e, system-averaged) coordination number $\bar{z}$ and the mean orientational order parameters ${\bar{\Phi}_4}$ and ${\bar{\Phi}_6}$, which provide information about the structural arrangement around the particles. For an individual particle $i$, the instantaneous coordination number $z_i$ is defined as the number of nearest neighbours at a given time $t$. Two particles are considered as neighbours if their distance is smaller than $r_{cl}=1.15\sigma$. The mean coordination number of the system (at time $t$) then follows as

\begin{equation}
	\bar{z}=\frac{1}{N}\sum_{i=1}^{N}z_i.\label{eq:co_ord_no.}
\end{equation}

Further, the orientational order parameter is defined for each particle $k$ as \cite{halperin1978theory,steinhardt1983bond}
\begin{equation}
	\Phi_n^k=\frac{1}{z_k}\left|\sum_{l=1}^{z_k}e^{in\theta_{kl}} \right|,\label{eq:abop}
\end{equation}

where $z_k$ is the number of nearest neighbours of particle $k$ and $\theta_{kl}$ is the angle between the vector $\mathbf{r}_{kl}=\mathbf{r}_k-\mathbf{r}_l$ connecting the central particle under consideration (denoted by index $k$) and one of its neighbours (denoted by index $l$), and the $x-$axis. We here consider, in particular, the cases $n=4$ and $n=6$, where $n=4$ refers to quadratic order and $n=6$ refers to hexagonal order. These cases are the most relevant ones for the 2D particle system studied here. The value of $\Phi_n^k$ is equal to 1 if the order is perfect in the sense that the particle $k$ has exactly $n$ neighbours and each each pair $(k,l)$ forms an angle of $\theta_{kl}=2\pi/n$ with the adjacent pair. In a realistic (fluctuating) configuration, a particle belonging to a quadratic structure is expected to have a value of $\Phi^k_4$ close to 1 and $\Phi^k_6$ close to 0 and vice-versa for a particle belonging to a hexagonal structure. The mean (system-averaged) orientational order parameter of the system is then defined as

\begin{equation}
	\bar{\Phi}_n=\frac{1}{N}\sum_{k=1}^{N}\Phi_n^k. \label{eq:mean_abop}
\end{equation}

At the low temperature ($T^{*}=0.05$) considered here, the structural parameters can significantly deviate from zero. The actual values of $\bar{z}$, $\bar{\Phi}_4$ and $\bar{\Phi}_6$ after a simulation run of $100\tau_B$ depend not only on the thermodynamic state defined by $T^*$ and $\rho^*$ but, in particular, on $\delta_r$. This is understandable, to some extent, from the properties of the dipolar potential (as function of the polar angle of the connection vector) shown in Fig.~\ref{fig:BD_model}(b).

At small values of $\delta_r$ (i.e., all charges located close to the centre), the attractive interactions are only weakly anisotropic; i.e., two particles experience essentially the same degree of mutual attraction regardless of their mutual arrangement. In this case, hexagon-like (and, thus, essentially isotropic) structures are generally preferred due to their packing efficiency. As a consequence, $\bar{z}$ assumes rather large values (as compared to its 2D maximum, $\bar{z}_{\text{max}}=6$), and the order parameter $\bar{\Phi}_6$ is large as well. With increasing $\delta_r$ the degree of anisotropy becomes larger and larger, as reflected by the more pronounced minima [see Fig.~\ref{fig:BD_model}(b)]. This eventually leads to the preference of quadratic structures where the connection vectors between two particles align along the direction of the induced dipole moments. Here, the mean coordination number is smaller (as compared to the hexagonal) case, and the $\bar{\Phi}_4$ order parameter dominates. Interestingly, as shown in Ref.~\cite{kogler2015generic} (and confirmed by the simulations in the present study), there is a sharp change between these structures when one varies $\delta_r$ at fixed (low) temperature and fixed (moderate) density. For the case $T^{*}=0.05$ and $\rho^{*}=0.3$, this change occurs at $\delta_r\approx 0.21\sigma$. 

In the remainder of this paper, we focus on the aggregation behaviour at three values of $\delta_r$, that is, $\delta_r=0.1\sigma$ (henceforth referred to as case~1), $ 0.21\sigma$ (case~2), and $0.3\sigma$ (case~3). The corresponding structural parameters after a simulation run of $100\tau_B$ are summarized  in Table~\ref{state_points}. Furthermore, each case is illustrated by a simulation snapshot shown in Figs.~\ref{fig:BD_model}(b)-(d). As expected from the previous discussion, case~1 and case~2 correspond to predominantly hexagonal and quadratic structures, respectively. Case~2 with $\delta_r=0.21\sigma$ is a boundary case where the two structures compete.

We note that all of these low-temperature aggregated structures are long-lived but {\em transient} in character. That is, on the time scale of the BD simulation, structural order parameters (defined above) assume quasi-stationary values, whereas certain time correlation functions such as bond correlations do not decay to zero (within the considered simulation time). We also note that it is still an open question whether the present model systems, with the patchyness considered here, exhibits a full (equilibrium) condensation transition if one would wait sufficiently long. This scenario (in which the clusters would be only by-products occurring within the two-phase regime) is indeed suggested based on mean-field-like arguments, but was not observed in the simulations~\cite{kogler2015generic}.

\begin{table} [htp!]
	\caption{Charge separation parameter, structural quantities, and classification for the three cases considered in this study. In all cases, $\rho^{*}=0.3$ and $T^{*}=0.05$. \label{state_points}}
	\begin{tabular}{|c|c|c|c|c|c|}
		\hline
		Case No. & $\delta_r$ & $\bar{\Phi}_4$ & $\bar{\Phi}_6$ & $\bar{z}$ & Overall structure  \\
		\hline
		1 & $0.1\sigma$ & 0.1705 & 0.9856 & 4.502 & Hexagonal  [Fig.~\ref{fig:BD_model}(b)]\\			
		2  & $0.21\sigma$ & 0.3194 & 0.8584 & 4.532 & Competing [Fig.~\ref{fig:BD_model}(c)] \\
		3  & $0.3\sigma$ & 0.9321 & 0.3071 & 2.728 &Quadratic [Fig.~\ref{fig:BD_model}(d)]\\
		\hline
	\end{tabular}
\end{table}

\section{Coarse-grained modelling}  \label{section: coarse_grained}
\subsection{Categorization} \label{sec: categorization}
\begin{figure} [htp!]
	\centering
	\includegraphics[width=0.5\textwidth]{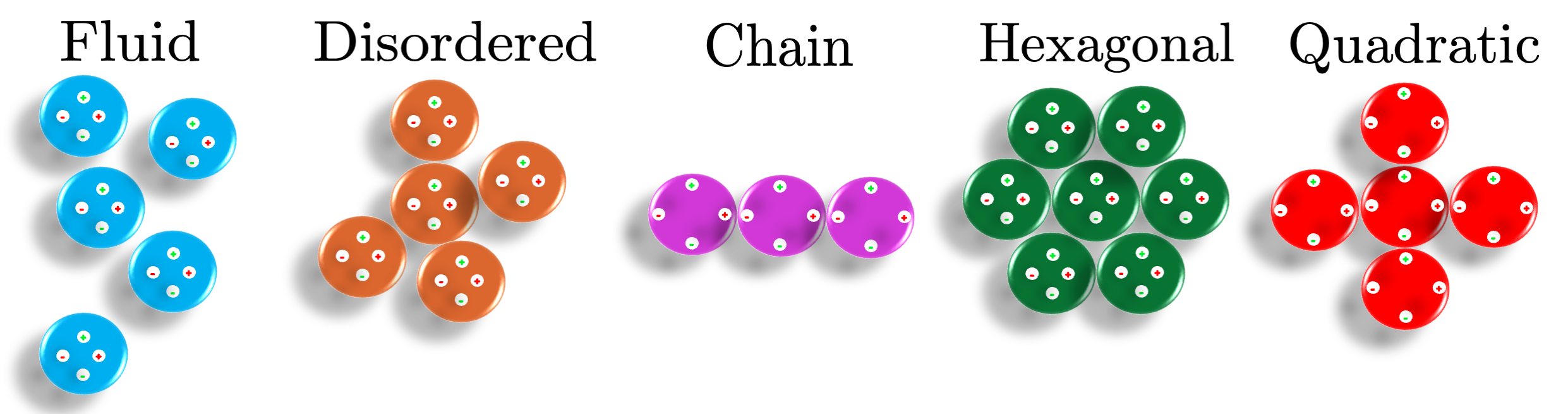}
	\caption{Schematic of the structural arrangement around a particle in each of the five states.} \label{structural_configs}
\end{figure}

We aim at investigating the system on a coarse-grained level where each particle is categorized into a discrete ''state'' according to the local structure in its immediate neighbourhood. This classification will later be the basis of our discrete state model. Specifically, we introduce five different states: fluid, chain-like, disordered, hexagonal, and quadratic, see Fig.~\ref{structural_configs} for an illustration.

\begin{figure*} [htp!]
	\centering
	\includegraphics[width=1.0\textwidth]{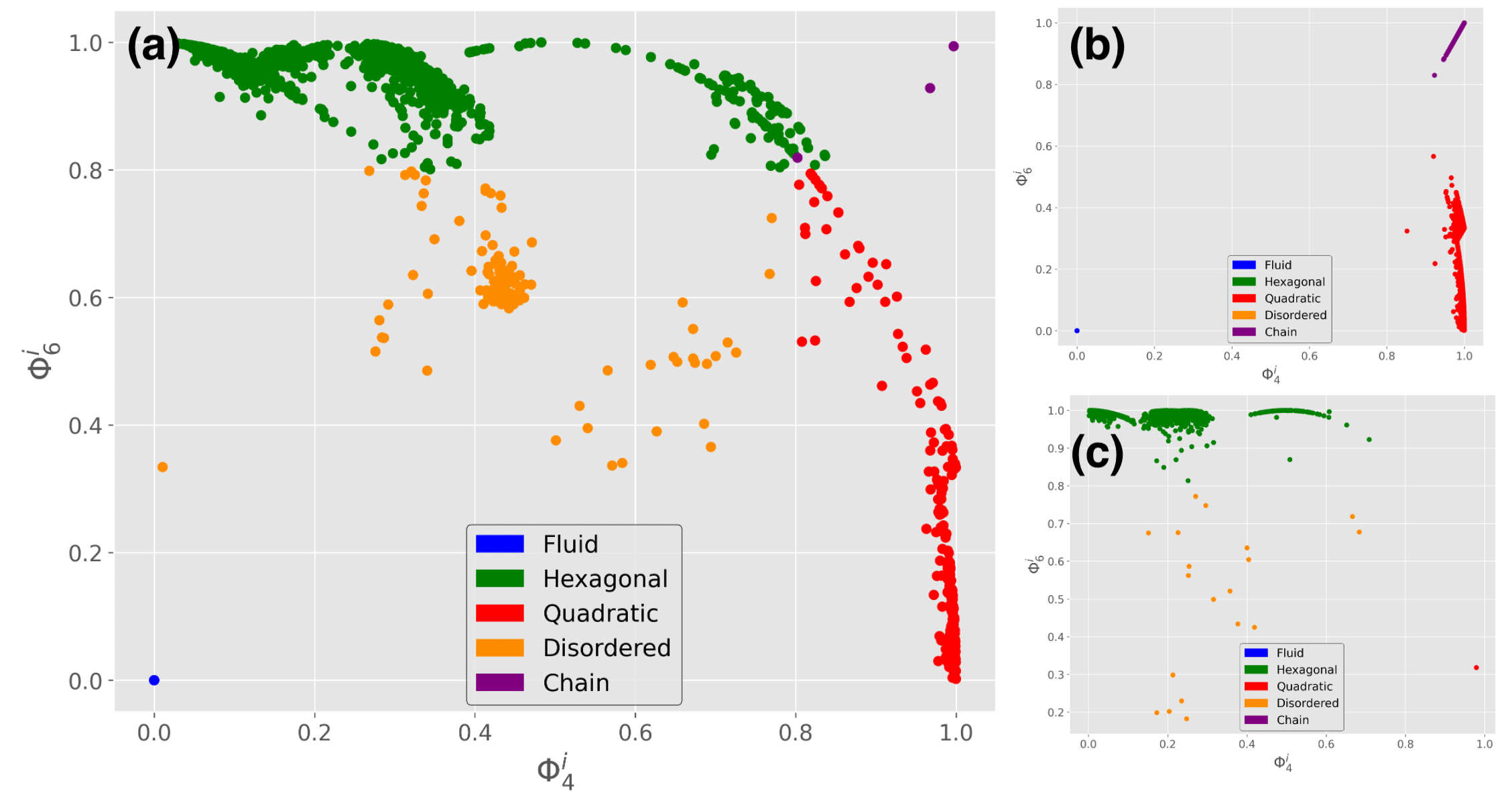}
	\caption{Scatter plot of the local orientational order parameters $\Phi_4^i$ and $\Phi_6^i$ of each particle $i$ after a simulation time of $t=100\tau_B$ for the system in (a) case~2 (b) case~3 and (c) case~1. The data points are colour coded according to the state that the particle was categorized into. } \label{fig:scatter}
\end{figure*}

The states are defined based on the (local) structural parameters introduced in Sec.~\ref{aggregates_studied}, that is, the co-ordination number $z_i$ and the orientational order parameters $\Phi_4^i$ and $\Phi_6^i$ [see Eq.~(\ref{eq:abop})] for each particle $i$. We start by checking whether the particle is in one of the highly ordered "states" (chain, hexagonal, or quadratic). This is the case if one, or both, of the parameters $\Phi_4^i$ and $\Phi_6^i$ take large values. Specifically, a particle is assigned to be "chain" if both $\Phi_6^i\geq0.8$ and $\Phi_4^i\geq0.8$. Indeed, it turns out that this condition is fulfilled only for particles with two neighbours oriented like a chain (see Fig.~\ref{structural_configs}). Further, we categorize a particle as belonging to the hexagonal (quadratic) state if only $\Phi_6^i\geq0.8$ ($\Phi_4^i\geq 0.8$), while the other parameter is smaller. In all other cases, we assign the particle's state according to the coordination number: "disordered" for $z_i\geq2$, "fluid" for $z_i\leq 2$. Performing many test calculations, we found that the chosen threshold values worked well even in situations where all of these different states appear [e.g., the system in case~2, Fig.~\ref{fig:BD_model}(c)]. To illustrate this, in Fig.~\ref{fig:scatter}, we show a scatter plot of the local orientational order parameters $\Phi_4^i$ and $\Phi_6^i$ of each particle $i$ for the three cases at $t=100\tau_B$. The data points are colour-coded according to the state that the particle was categorized into. For cases 1 and 3 [Figs.~\ref{fig:scatter}(c) \& (b)] the categorization is rather straightforward since most of the particles are forming hexagonal (in case~1) and quadratic/chain-like structures (in case~3). In contrast, case~2 is more delicate [see Fig.~\ref{fig:scatter}(a)], due to the competition between the structures. Still, as shown in the corresponding scatter plot [Fig.~\ref{fig:scatter}(a)], our categorization scheme is able to successfully categorize particles into the five different states. In particular, particles with high $\Phi_4^i$ ($\Phi_6^i$) and low $\Phi_6^i$ ($\Phi_4^i$) are categorized correctly as being in the quadratic (hexagonal) state (according to the colour-code in Fig.~\ref{fig:scatter}). Particles belonging to the disordered, fluid and chain states are also detected successfully. 

\begin{figure*} [htp!]
	\centering
	\includegraphics[width=1.0\textwidth]{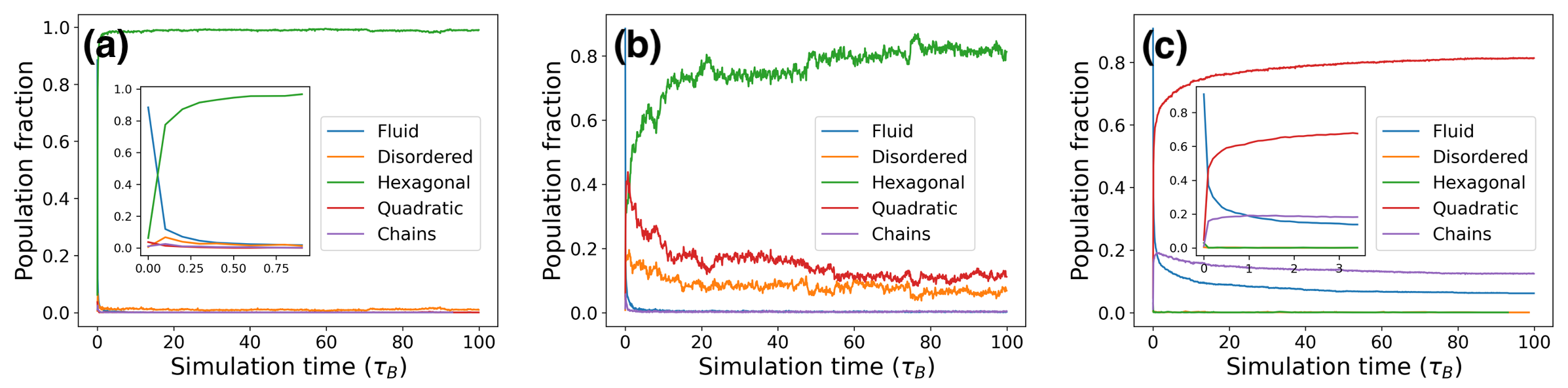}
	\caption{Population fractions observed in BD simulations as a function of time for (a) case 1, (b) case 2, and (c) case 3.} \label{fig:pop-t}
\end{figure*}

The categorization can be done, in principle, at each instant in time. As an example, we plot in Fig.~\ref{fig:pop-t} the time evolution of the population fraction of particles in each of the five states, for all the three cases, for one noise realization. The simulations are started from a random, uniform distribution of particles (almost all particles in the fluid state). For case~1 [Fig.~\ref{fig:pop-t}(a)] the population fraction of the hexagonal state shoots up almost immediately to a value very close to unity (at around $t\sim$1.0$\tau_B$). This suggests that all the fluid particles directly aggregate into the hexagonal state. The population fraction of all the other states are observed to be close to zero. This behaviour can be attributed to the strong preference of hexagonal structures in case~1 (see Section~\ref{aggregates_studied}). In case~2 [Fig.~\ref{fig:pop-t}(b)], the particles in the fluid state immediately form aggregates in hexagonal, quadratic and disordered states. This is due to the the lack of a preferred structural configuration in case~2 (see Section~\ref{aggregates_studied}). Although the system consists of an equal proportion of hexagonal and quadratic structures at initial times (around $t\sim5\tau_B$), only hexagonal structures continue to grow, saturating at a value of around $0.8$. In contrast, the population fraction of the quadratic structure decreases and saturates to a value of around $0.15$. The fraction of disordered particles remains at a value of around $0.1$ throughout the simulation. We also find that the population fractions are observed to fluctuate more strongly for case~2 compared to the other two cases. Finally, for case~3 [Fig.~\ref{fig:pop-t}(c)], the fluid particles transition directly into the quadratic and chain states. However, we do not see the fluid-state population fraction go to zero, unlike cases 1 and 2. Rather, it stagnates at a value of around $0.1$. This is because particles at the edges of the chains (with only a single nearest neighbour) are categorized as belonging to the fluid state. At around $t\sim20\tau_B$, the system is dominated by quadratic structures amounting to a population fraction of around $0.75$ and chain-like structures (population fraction of around $0.19$). Further fluctuations in the population fractions are not observed. This can be attributed to the strong preference of particles to be aligned along the $x-$ and $y-$ axes for the parameter combination in case~3 (see Section~\ref{aggregates_studied}).

\subsection{Discrete states dynamics} \label{MSM_theory}
Our next task is to describe the dynamics of the system's local structural configurations as a series of stochastic, memoryless jumps between the set of discrete states defined in Section \ref{sec: categorization}. Inspired by the theory of Markov state models (MSM) \cite{bowman2013introduction,prinz2011markov,husic2018markov}, we assume that we have identified $m$ discrete Markov states. The population probabilities of each state at time $t$ can then be expressed as a $m-$component vector $\mathbf{P}(t)$ with each element $P_{\alpha}(t)$ (with $\alpha=1,...,m$) representing the probability that a particle chosen at random from the system exists in state $\alpha$. We then define the $m\times m$ transition probability matrix $\mathbf{T}(\tau)$. The elements of $\mathbf{T}(\tau)$ consist of the conditional probabilities of a particle transitioning from a state $\alpha$ to state $\beta=1,...,m$  after a time $\tau$, given that the particle was initially in state $\alpha$. The timescale $\tau$ is called the lag-time. It is chosen such that the dynamics is Markovian beyond $\tau$. Once $\mathbf{T}(\tau)$ and $\mathbf{P}(t)$ are known, the population probability vector after time $\tau$, $\mathbf{P}(t+\tau)$ can be obtained as \cite{bowman2013introduction,prinz2011markov}
	
\begin{equation}
	\mathbf{P}(t+\tau)=\mathbf{T}(\tau)\mathbf{P}(t).   \label{eq:CK}
\end{equation}
	
This equation is known as the Chapman-Kolmogorov equation for discrete processes, and represents an alternative representation of the Master equation \cite{buchete2008coarse,kube2007coarse,metzner2007generator,schutte2011markov,sriraman2005coarse} (see, e.g., Ref. \cite{prinz2011markov}). 
	
For stationary processes, upon performing successive operations, the population probability vector at later points in time (after each time interval $\tau$) can be computed using only the population probabilities at the arbitrary initial time $t$, that is, 
	
\begin{equation}
	\mathbf{P}(t+s\tau)=\mathbf{T}^s(\tau) \mathbf{P}(t), \label{eq:MSM_evo}
\end{equation}

where $s$ is an integer. Because of Eq.~(\ref{eq:CK}), for stationary processes, the left hand side of Eq.~(\ref{eq:MSM_evo}) can also be written as $\mathbf{P}(t+s\tau)=\mathbf{T}(s\tau)\mathbf{P}(t)$. It thus follows that a transition probability matrix evaluated at a lag-time of $s\tau$ represented by $\mathbf{T}(s\tau)$, is equivalent to $\mathbf{T}^s(\tau)$.

\subsection{Transition probabilities} \label{construction}
As a first step towards the construction of the transition probability matrix $\mathbf{T}(\tau)$, we evaluate the so-called transition count matrix $\mathbf{C}(\tau)$. Specifically, the elements $C_{\alpha\beta}$ of $\mathbf{C}(\tau)$ count the number of particles that transition from state $\alpha$ to state $\beta$ after the lag-time $\tau$. On dividing the number of transitions by the total number of transitions originating from that particular state, we obtain the transition probability matrix $\mathbf{T}(\tau)$ with elements \cite{bowman2013introduction,prinz2011markov}
\begin{equation}
	T_{\alpha \beta}(\tau)= \frac{C_{\alpha \beta}(\tau)}{\sum_{\gamma} C_{\alpha \gamma}(\tau)}. \label{eq:TPM}
\end{equation}

At this point it is worth to briefly mention one important property of the matrix $\mathbf{C}(\tau)$ assumed in MSM descriptions. One typically assumes $\mathbf{C}(\tau)$ to be symmetric, expressing the fact that every time there is a transition from state $\alpha$ to state $\beta$, there is also a transition from state $\beta$ to $\alpha$, that is, the so-called detailed balance condition is fulfilled. To actually enforce this condition between each pair of states, $\mathbf{C}(\tau)$ is usually symmetrized in the following way (before $\mathbf{T}(\tau)$ is computed) \cite{bowman2013introduction}
	
\begin{equation}
	\hat{C}_{\alpha \beta}(\tau)=\frac{C_{ \alpha \beta}(\tau)+C_{ \beta \alpha}(\tau)}{2}. \label{eq:TCM_symmetrization}
\end{equation}
	
This symmetrization is justified if the system is truly in thermodynamic equilibrium (and thus, detailed balance is fulfilled) for the time scale considered. However, colloidal self-aggregation is a time-dependent process \cite{krinninger2016minimal,perkett2014using}, such that detailed balance cannot be taken for granted. We will come back to this point in Section \ref{detailed_balance}.

\subsection{Decomposition}  \label{tpm_properties}
Having obtained the matrix $\mathbf{T}(\tau)$, we calculate its eigenvalues and eigenvectors to extract information regarding population probabilities. In classical MSM theory, the underlying assumptions are\cite{prinz2011markov,bowman2013introduction} that the system is in thermodynamic equilibrium and is, thus, ergodic (i.e., each of the discrete states is accessible from another one) and stationary (i.e., the same equilibrium distribution is reached irrespective of the initial distribution). It can be shown\cite{prinz2011markov} that under these conditions, the eigenvalues and eigenvectors of $\mathbf{T}(\tau)$  have the following properties. The largest eigenvalue is 1 with the remaining eigenvalues being smaller than 1, i.e., $1>\lambda_2>\lambda_3>\lambda_4>....\lambda_m$. The normalized eigenvector corresponding to the largest eigenvalue $\lambda_1$ represents the state population probability of each of the (pre-defined) Markov states in the stationary state.

\subsection{Handling non-stationarity}    \label{handl._stationarity}
As described in Sec. \ref{sec: categorization}, the discrete state model developed here targets the dynamics of local structural changes rather than global ones. During the aggregation process, the dominant type of these changes and thus, the transition probability matrix, $\mathbf{T}(\tau)$ can vary in time until the system has reached a global equilibrium state (i.e., the minimum of the free energy). Therefore all quantities characterizing the discrete state model, particularly the transition probability matrix, can, in principle, be non-stationary during the aggregation process. 
	
For example, we observe significant changes in $\mathbf{T}(\tau)$ when two clusters collide and merge to form a larger cluster. In this situation, the elements of $\mathbf{T}(\tau)$ are found to relax into different "quasi-stationary" values before (and after) each collision. There is also a significant timescale separation between the cluster growth timescales and the timescale for the transition of particles between different local structures. Specifically, in our simulations, local structural transitions were observed to occur at timescales at or below $t_{\text{local}}=10^{-3}\tau_B$. At the early stages of aggregation (when the largest cluster size is below 25 particles), the largest cluster in the system was observed to grow at timescales \textit{at least} two times longer than $t_{\text{local}}$, $t_{\text{largest}}\simeq2\times t_{\text{local}}$. At later stages of the aggregation, $t_{\text{largest}}$ was observed to reach as high as $30\tau_B$ (i.e. $t_{\text{largest}}\simeq10^{4} \times t_{\text{local}}$). Motivated by these observations, we monitor the number of particles in the largest cluster, $n$, and work under the assumption that $\mathbf{T}(\tau)$ relaxes into a quasi-stationary state after each change in $n$. In other words, we discretize the system based on the value of $n$. Similar parameters have also been utilized in studies of the crystallization of Brownian particles under shear\cite{lander2013crystallization}, self-assembly of dodecahedral capsids from pentagonal subunits \cite{trubiano2024markov} and that of $T=3$ capsids from triangular subunits \cite{trubiano2024markov}. To take into account transitions occurring simultaneously in different clusters (not just in the largest one), we apply the independent Markov decomposition (IMD) assumption \cite{hempel2021independent}. This is justified since, at the rather low density considered in our simulations, multiple clusters form in different, spatially separated regions [see Fig.~\ref{fig:BD_model}(c)]. Hence, each cluster can be treated as a quasi-independent subsystem, consistent with the IMD assumption. Local transitions in all these clusters are then used to construct the transition probability matrix.
	
\begin{figure} [htp!]
	\centering
	\includegraphics[width=0.5\textwidth]{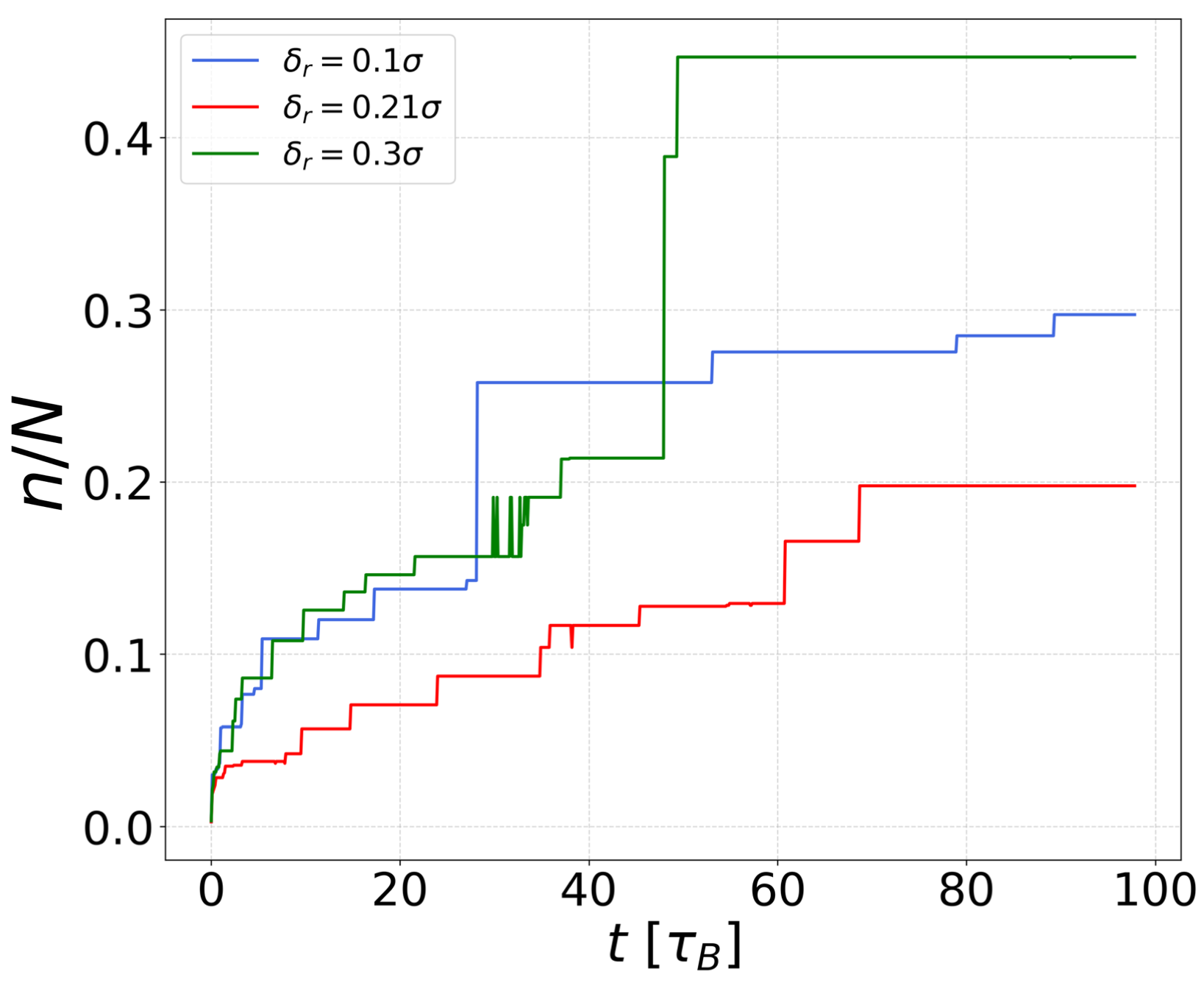}
	\caption{Fraction of particles in the largest cluster, $n/N$, as a function of time $t$ for a single noise realization of the three cases studied here (see Table \ref{state_points}).} \label{fig:n-t_coexistence}
\end{figure}

To calculate $n$ from the Brownian dynamics simulations, we use the same distance criterion already introduced for the definition of nearest neighbours (see Sec. \ref{aggregates_studied}). Particles whose centre-to-centre distances are smaller than $r_{cl}=1.15\sigma$ are defined as being "bonded", and all particles that are bonded with each other are defined as belonging to the same cluster. In Fig.~\ref{fig:n-t_coexistence}, we present our numerical results for the quantity $n/N$ as a function of time for all four cases listed in Table \ref{state_points}. The data are obtained for one particular noise realisation. However, the general trends observed here were found to be robust for different noise realisations. 
	
It can be seen that, in general, $n/N$ increases in a step-like manner staying at a particular value for a certain period of time before increasing again. During this time period, $\mathbf{T}(\tau)$ relaxes into a quasi-stationary state. Based on this behaviour, we construct the discrete state model in the following way: after each change of the value of $n$, we compute a new transition probability matrix, $\mathbf{T}_n(\tau)$, each one corresponding to a different quasi-stationary state within the aggregation process. Ultimately, in this manner, we can utilize the parameter $n$ to handle the time dependent nature of the process. We would also like to mention that calculating implied timescales in the standard way (i.e. by monitoring eigenvalues of the transition probability matrix as functions of lag times) is not straightforward for our MSM. We obtain a new set of implied time-scales for each of the $\mathbf{T}_n(\tau)$ at different values of $n$. Moreover, the timescales were  not observed to properly saturate, making physically interpreting these results challenging.

From a physical point of view, we can understand the behaviour of $n/N$ in the different cases shown in Fig.~\ref{fig:n-t_coexistence} as follows. For case~2 , $n/N$ increases in relatively small steps approaching a value of around $n/N\simeq0.2$ after $100\tau_B$. In contrast, in case 1 and 3, $n/N$ shows a large jump at a certain time, a behaviour not seen for case 2. To understand this difference, we note that the value $\delta_r=0.21\sigma$ chosen in case 2 allows for the co-existence of hexagonal and quadratic structure (as explained in Section \ref{aggregates_studied}). In other words, there is no clear preference of any structure; rather, the particles are allowed to re-orient and bond with other clusters. This ultimately leads to quite compact clusters in case 2 [see Fig.~\ref{fig:BD_model}(c)]. In contrast, for $\delta_r>0.21\sigma$ and $\delta_r<0.21\sigma$ (as in cases 1 and 3), there is a preferred cluster structure (hexagonal or quadratic), resulting in rigid bonds between the particles. At the same time, the clusters tend to be more spread out in space as compared to case 2. This, in turn, increases the likelihood of cluster merges, causing the value of $n/N$ to increase rapidly within a short period of time.

\section{Predictions from the coarse-grained model}  \label{results}
\subsection{Sampling Details}   \label{sampling}
In the present study, we have calculated the matrices $\mathbf{T}_n(\tau)$, based on 50 noise realisation of the particle-resolved dynamics with $N=1800$. The particle positions are initialised randomly (yet avoiding overlaps) such that most particles at $t=0$ can be categorized as belonging to the fluid state. The lag-time for the sampling is chosen as $\tau=100\Delta t=10^{-3}\tau_B$, i.e., each system configuration (positional data of all particles in the system) that have a time different of $\tau$ between them are used to construct the transition probability matrices. The particular value of $\tau$ is chosen such that it is small enough for the proper discretization of $n$ (especially at the early stages of aggregation) and the observation of transitions, yet, large enough to sample long BD simulations in a computationally efficient manner. However, as stated in Section \ref{handl._stationarity}, new transition probability matrices $\mathbf{T}_n(\tau)$ are obtained after each change in $n$. Since $\mathbf{T}_n(\tau)$ was also found to be dependent on the model parameters ($\rho^*$, $T^*$ and $\delta_r$), we calculate $\mathbf{T}_n(\tau)$ separately for each of the different cases studied here. 

\subsection{Model validation}  \label{validation}
\begin{figure*} [htp!]
	\centering
	\includegraphics[width=1.0\textwidth]{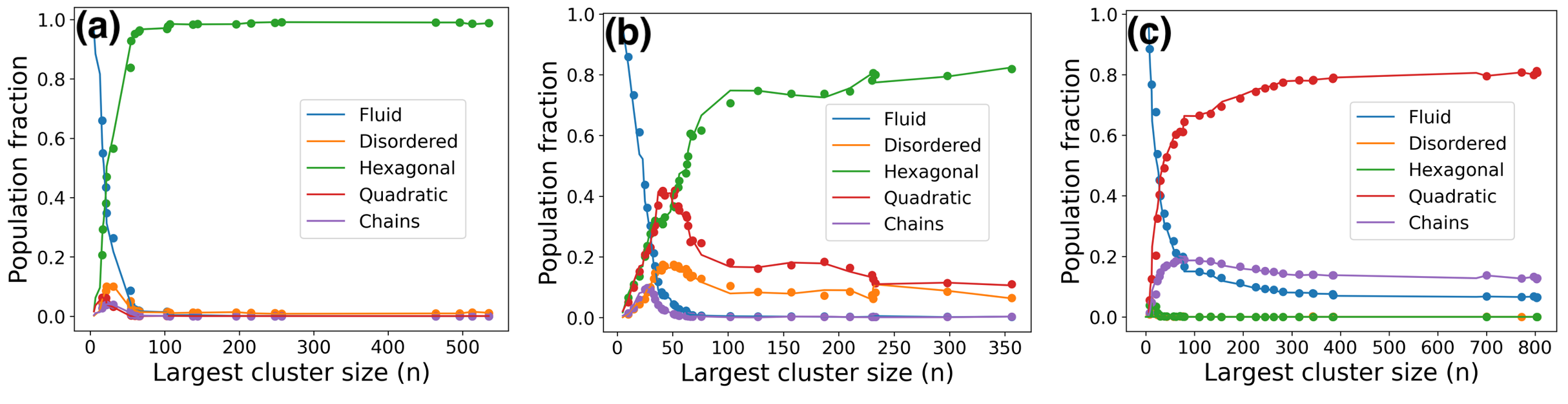}
	\caption{Population fractions observed in particle-resolved simulations (lines) and stationary population probabilities predicted by the discrete state model (dots) as a function of $n$ in a single noise realisation for (a) case 1, (b) case 2, and (c) case 3.} \label{fig:CK_sym}
\end{figure*}

In order to validate the discrete state model, we compare the elements of the normalised, five-dimensional eigenvector corresponding to the largest eigenvalue (that is, the predicted stationary population fractions), with the population fractions of the five discrete states calculated in the particle-resolved simulations (according the the state discretization in Section \ref{sec: categorization}). This is done at different values of $n$ for a single noise realisation. As seen in Fig.~\ref{fig:n-t_coexistence}, the system stays at a value of $n$ for a certain time-window. Hence, for comparison, we choose population fractions from the particle-resolved simulations at the time point right before the next change in $n$ occurs. We expect agreement between these two quantities if the discrete states are chosen such that they satisfy the conditions described in Section~\ref{tpm_properties} and the transitions occur within the time-scale in that we sample $\mathbf{T}_n( \tau)$ \cite{bowman2013introduction}. The results are shown in Fig.~\ref{fig:CK_sym} for the three cases considered here. Indeed, it can be seen that the predictions match quite well with the simulated population fractions for all the cases studied. It is worth to mention that, in other recent works on MSM models for self-aggregation \cite{trubiano2024markov, trubiano2022optimization, trubiano2021thermodynamic, perkett2014using}, finer discretization schemes have been used. This can lead to a significantly larger number of discrete states compared to the five states in the present study. Such a finer discretization could, in principle, also be used for our system. For example, one could differ between particles in different sized clusters or particles in the interior or at the boundaries of the cluster. Monitoring the corresponding population fractions could provide additional information on the aggregation process. However, in our view, such a fine discretization scheme is necessary only when prior information about quasi-stationary states is not known \cite{bowman2013introduction}. For the present system, we have clear information regarding occurring structures and the discrete states are chosen accordingly (see Section~\ref{sec: categorization}). 

We proceed with a brief description of the trends seen in the population fractions as a function of $n$ for each of the cases in Fig.~\ref{fig:CK_sym}. Qualitatively, the trends are similar to those of the population fraction as a function of time (Fig.~\ref{fig:pop-t}). However, in the early aggregation stages (corresponding to $n\leq50$), $n$ can evolve in time-scales shorter than the ones used for the plots in Fig.~\ref{fig:pop-t}. So, plotting the population fractions as a function of $n$ is a convenient way to resolve population dynamics in the early cluster aggregation stages. For the population dynamics of case 1 [see Fig.~\ref{fig:CK_sym}(a)], at $n\sim25$, we see, in addition to the formation of hexagonal aggregates [observed in the population fractions vs. time plot in Fig.~\ref{fig:pop-t}(a)], the formation of a small fraction of disordered, quadratic and chain-like aggregates (each state's population fraction amounting to slightly less than $0.1$). These eventually convert into the hexagonal state as $n$ grows. In case~2 [see Fig.~\ref{fig:CK_sym}(b)], in addition to the growth of the fraction of particles in the hexagonal, quadratic and disordered states [seen in the population fractions vs. time plot in Fig.~\ref{fig:pop-t}(b)] we are also able to resolve an increase of the chain state population fraction (at $n\sim25$). Finally, for case~3 [Fig.~\ref{fig:CK_sym}(c)], we do not observe the increase of any state's population fraction that was not observed in Fig.~\ref{fig:pop-t}(c), but we are able to better resolve the dynamics by obtaining the plot as a function of $n$.

\subsection{Predicting population fractions as functions of time}  \label{sec: pred_as_functionof_time}
So far, we have focused on predicting population fractions as functions of $n$. However, despite the advantages (resolution at early stages), we are also interested in predicting population fractions as functions of time. An obvious advantage to the latter functions is that these quantities can be directly measured (or calculated) from particle-resolved simulations or experiments. Further, these data are more natural for the description of time-dependent processes. Finally, time-dependent population fractions are more suitable for noise-averages. Indeed, we observed the calculation of noise-averaged population fractions as functions of $n$ to be challenging (note that the data in Fig.~\ref{fig:CK_sym} are for one single noise realization). This is because obtaining consistent, repeating values of $n$ in different noise realisations is often not possible. Certain values of $n$ can be missed entirely and most values are rarely repeated.

In this section, we therefore extend our methodology to predict noise-averaged or single-realization population fractions as \textit{functions of time}. Although, on the particle level, the calculation is straightforward (see Fig.~\ref{fig:pop-t}), obtaining the transition probability matrices in the coarse-grained model requires additional steps.

\subsubsection{Inferring the continuous $\mathbf{T}_n(\tau)$ vs. $n$ function}   \label{section: function_fit}
\begin{figure} [htp!]
	\centering
	\includegraphics[width=0.25\textwidth]{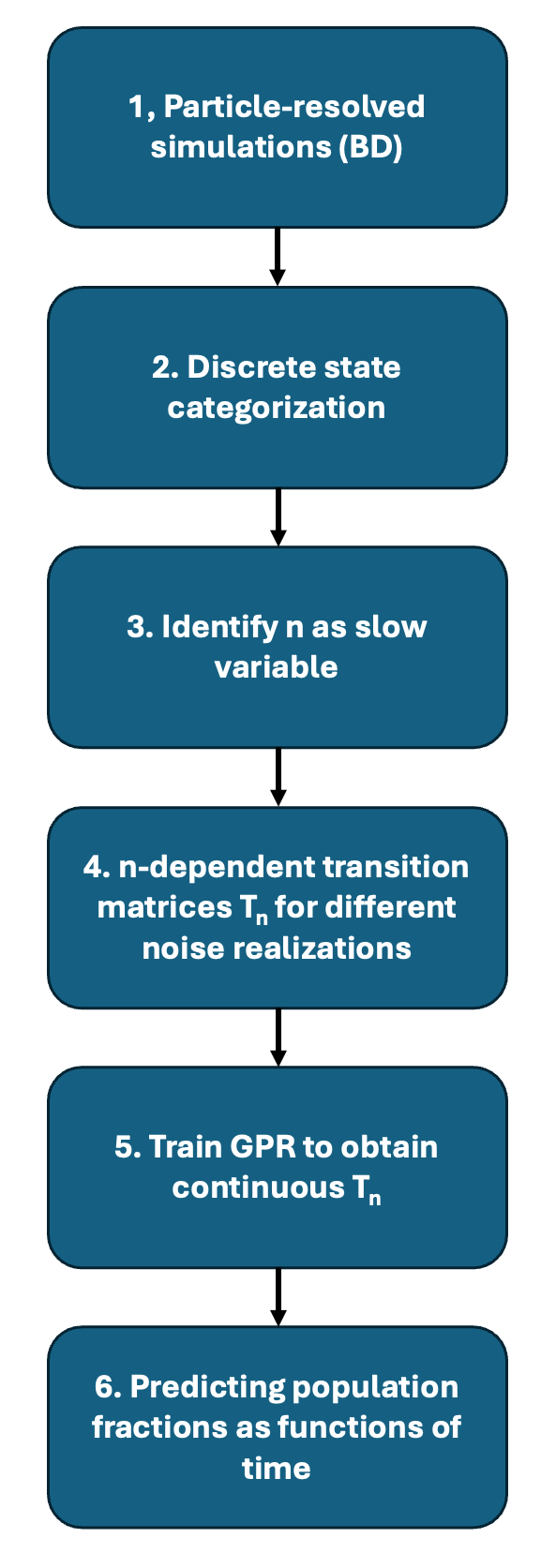}
	\caption{Flow diagram showing the steps performed to obtain a trained Gaussian parametric model for predicting transitions and population fractions at any given time-point from particle-resolved simulations.} \label{fig:diagram_flow}
\end{figure}

As a first step towards predicting population fractions as a function of time, we have to obtain a continuous function that maps $\mathbf{T}_n(\tau)$ to $n$. So far, since the largest cluster in our system grows in step-like increments (see Fig.~\ref{fig:n-t_coexistence}), we only have a discontinuous function of $\mathbf{T}_n(\tau)$ vs. $n$. Even after performing simulations of 50 different noise realisations, we have observed that certain values of $n$ can remain unobserved. Here, we suggest the following solution: since we expect $\mathbf{T}_n(\tau)$ to vary only little between nearby values of $n$ (see Section \ref{handl._stationarity}), we can interpolate between existing data points to obtain $\mathbf{T}_n(\tau)$ at the missing values of $n$. To this end, we use a non-linear, non-parametric regression technique, namely, Gaussian process regression (GPR) \cite{williams2006gaussian, deringer2021gaussian}. Using GPR also allows us to use the different noise realisations as additional training data. A short summary of the technique and the kernels used for GPR in this study is given in Appendix \ref{app_norm_udip}. We use the GPR implementation in the Python package scikit-learn \cite{scikit-learn}. For performing GPR, we use $n$ and $\mathbf{T}_n(\tau)$ calculated from 50 different noise realisations for the three different  cases separately (see Section~\ref{sampling}). Once the GPR model is trained, it takes a value of $n$ as input (including unobserved and non-integer values) and estimates the corresponding transition matrix. In Fig.~\ref{fig:diagram_flow}, we represent the steps taken to fit the GPR model in a flow diagram. 

\subsubsection{Noise-averaged population fraction predictions}   \label{sec: noise_avg_preds}
In this section, we explicitly show that population fractions at any given time point $t$ can be predicted accurately using the trained GPR model (Section~\ref{section: function_fit}) if we have the size of the largest cluster at $t$. We here specifically focus on predicting the dynamics of the \textit{noise-averaged} population fractions as functions of time. This specific task is chosen for two main reasons. First, it acts as a good test for the accuracy of the transition matrices estimated by the GPR model. As an input, we need the the average (over the different noise realizations) size of the largest cluster as a function of $t$. These input values will include data that was not used for the training and optimization of the GPR, including non-integer values (as a result of the averaging). Second, the noise-averaged population fraction dynamics are quantities that can be reproduced consistently, as opposed to single noise-realizations. 

\begin{figure} [htp!]
	\centering
	\includegraphics[width=0.5\textwidth]{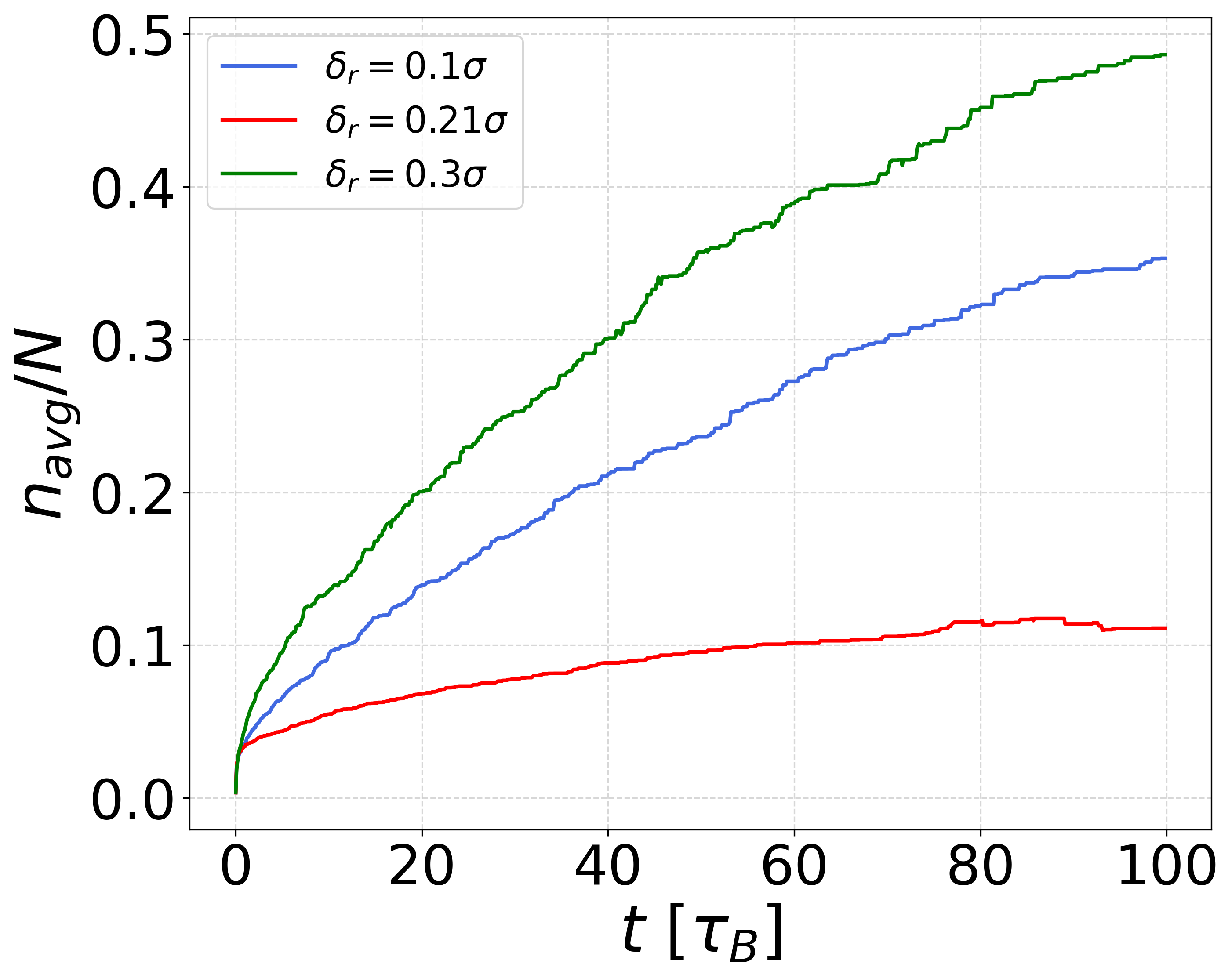}
	\caption{Fraction of particles in the largest cluster, $n_{avg}/N$, as a function of time, averaged over 50 noise realization of the three cases studied here (see Table \ref{state_points}).} \label{fig:n-t_avg}
\end{figure}

In order to infer the right transition probability matrix at each $t$ for each case (noise-averaged over 50 realizations), we first calculate the average size of the largest cluster [$n_{avg}(t)$] as a function of $t$. As shown in Fig.~\ref{fig:n-t_avg}, the step-like increase of $n$ as a function of time that was observed for a single noise-realisation (Fig.~\ref{fig:n-t_coexistence}) is no longer observed when we plot the quantity $n_{avg}$ (averaged over 50 noise realisations) as a function of time. We rather see, as expected, that $n_{avg}(t)$ evolves in a more continuous manner.

\begin{figure*} [htp!]
	\centering
	\includegraphics[width=1.0\textwidth]{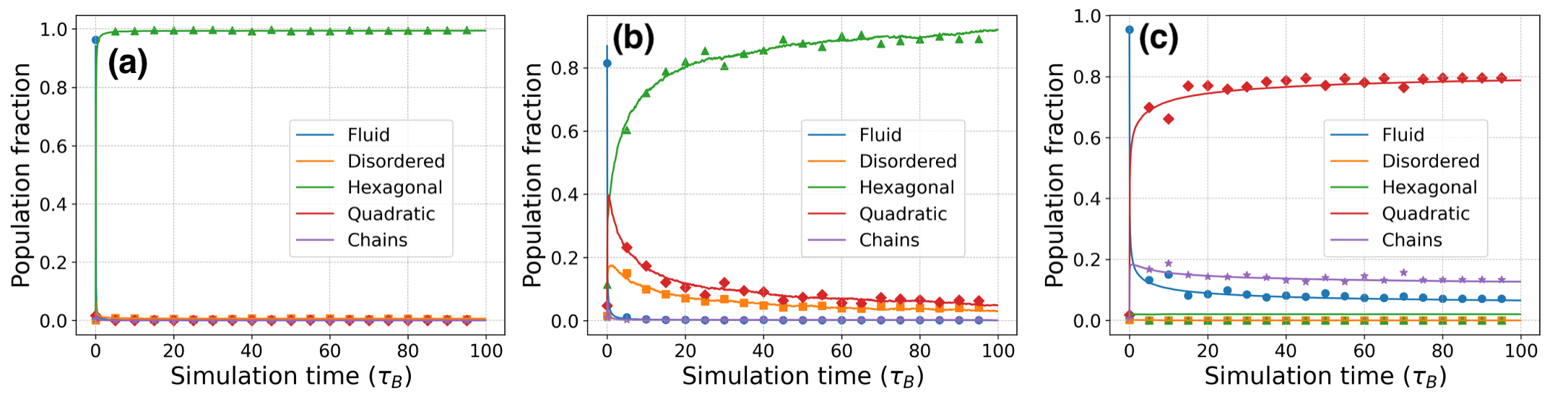}
	\caption{Noise-averaged population fractions observed in particle-resolved simulations (lines) and stationary population probabilities predicted by the discrete state model (symbols) as a function of time $t$ for (a) case 1, (b) case 2, and (c) case 3.} \label{fig:pop_t_preds}
\end{figure*}

Still, consistent with the trends discussed before (see Section~\ref{handl._stationarity}), the most rapid increase of $n_{avg}(t)$ is observed for case~3 and the slowest increase for case~2. We now pass the values of $n_{avg}(t)$ at each $t$ to the GPR model to estimate the corresponding transition probability matrices. Similar to the model validation performed in Section~\ref{validation}, the normalised, five-dimensional eigenvector of $\mathbf{T}_{n_{avg}}(\tau)$, then gives us the predicted population fractions at the corresponding $t$. This way, we can predict the population fractions at new, unobserved $n$ values without performing any additional simulations. In Fig.~\ref{fig:pop_t_preds}, the noise-averaged population fractions of the five states calculated from the particle-resolved simulations (averaged over 50 noise realisations, see Section~\ref{sampling}) to those predicted by the corresponding \textit{estimated} transition probability matrices $\mathbf{T}_{n_{avg}}(\tau)$, as a function of time. We see that the predictions of our discrete model agree quite well with the population fractions calculated from the particle simulations.

A limitation of the present approach is that the predictions are restricted within the range of finite $n$ values at which we have sampled the transition probability matrices and trained the GPR model with. Since we cannot access the long timescales required to reach high values of $n$ in our particle simulations, calculating transition probability matrices here is rather challenging. Although we can use Gaussian process regression to estimate (by extrapolation) the transition probability matrices at unobserved $n$ values beyond the range of the observed data, the accuracy of the estimations decrease, the larger the deviation is from the observed range. Recent works have started to explore different techniques to overcome such challenges like adaptive sampling and better discretization techniques of $n$ based on error analysis \cite{trubiano2024markov}. 

\subsubsection{Predictions after parameter switching}
In this section, we demonstrate that using our approach, we can predict population fraction dynamics for previously unseen scenarios without performing any additional BD simulations or sampling. Specifically, we study the case where, for a single noise realisation, the parameter $\delta_r$ is abruptly changed from $0.3\sigma$ to $0.1\sigma$ at a certain time, keeping all the other parameters the same (i.e., we change from case~3 to case~1). We specifically focus on this case because the aggregation mechanism changes from favouring quadratic to hexagonal aggregates. 

\begin{figure} [htp!]
	\centering
	\includegraphics[width=0.5\textwidth]{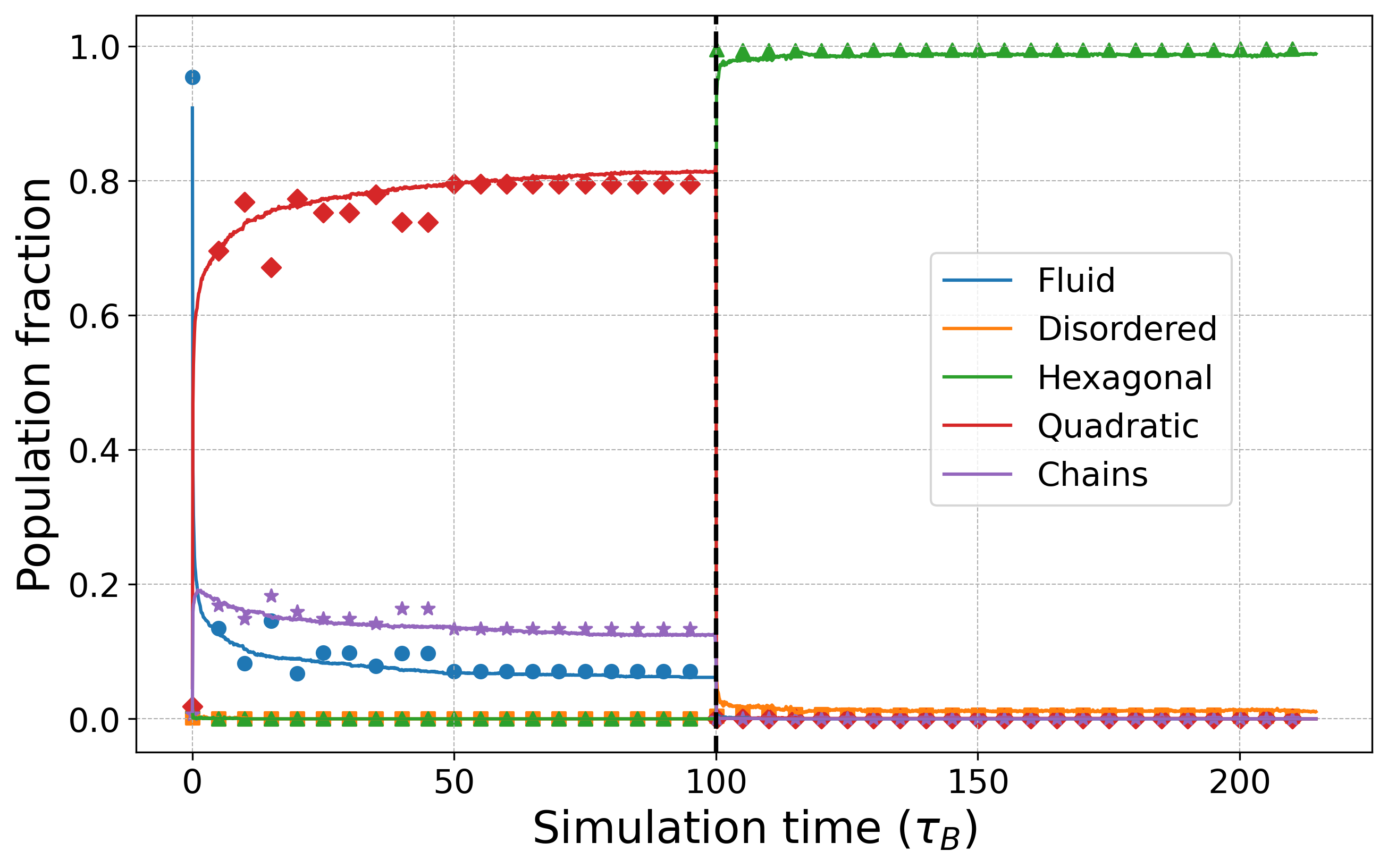}
	\caption{Population fractions observed in particle-resolved simulations (lines) and the predicted population fractions (symbols) as a function of time. The black, vertical dashed line at $t=100\tau_B$, represents the time at which the parameter is instantaneously changed from $\delta_r=0.3\sigma$ to $\delta_r=0.1\sigma$. } \label{fig:pop-t_change_preds}
\end{figure}

In order to compare the predictions to the population fractions from BD simulations, we perform a simulation with $\delta_r=0.3\sigma$ until $t=100\tau_B$ and then instantaneously change $\delta_r$ to $0.1\sigma$ and continue the simulation until $t=225\tau_B$. As was done in Section~\ref{sec: noise_avg_preds}, we use the trained GPR models (see Section~\ref{section: function_fit}) to estimate transition matrices for each time point. We use the models trained for case~3 for the predictions until $t=100\tau_B$ and that for case~1 for the predictions of the remaining time points. Similar to the predictions in Section \ref{sec: noise_avg_preds}, to find the right transition matrices for each time point $t$, we calculate the largest cluster size $n_t$ from the BD simulation at time $t$ and pass it on to respective GPR model to obtain the corresponding $\mathbf{T}_{n_t}(\tau)$. In Fig.~\ref{fig:pop-t_change_preds} we show the comparison between the predictions and the observed population fractions from the BD simulations as a function of time. It can be seen that the predictions are in good agreement with the population fractions observed in the BD simulation. This agreement is indeed quite remarkable, since the GPR models for both the cases used for the predictions here were trained from data that has been sampled from simulations without parameter change. 

\begin{figure*} [htp!]
	\centering
	\includegraphics[width=1.0\textwidth]{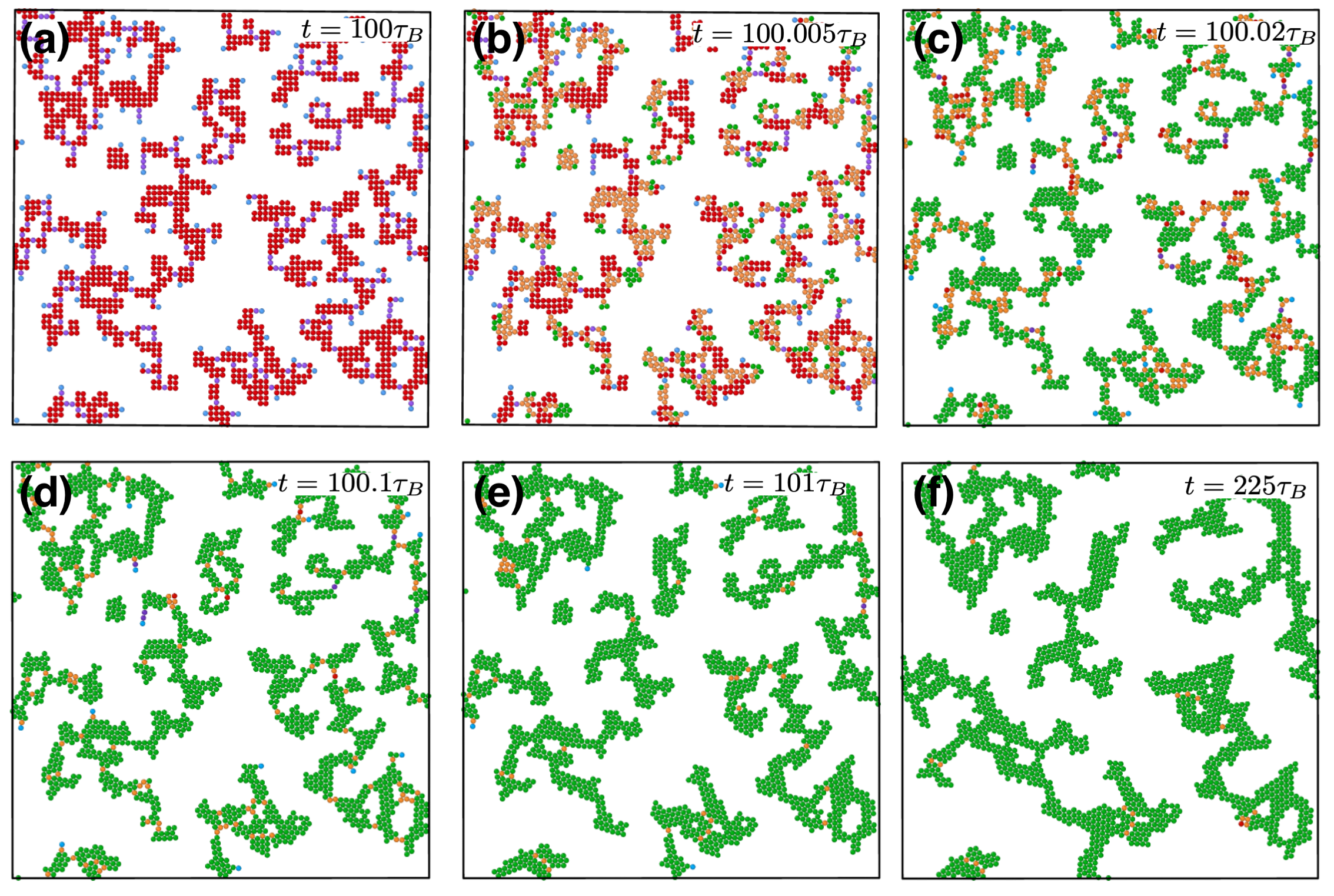}
	\caption{Simulation snapshots of the system undergoing the parameter change from case~3 to case~1. Figure (a) is at time $t=100\tau_B$, right before the parameter switch. Figures (b)-(f) are at different time points after the parameter switch with (b) $0.005\tau_B$, (c) $0.02\tau_B$, (d) $0.1\tau_B$, (e) $1\tau_B$ and (f) $125\tau_B$ after the parameter switch. The particles are colour coded according to the state they are categorized into: blue for fluid, green for hexagonal, red for quadratic, violet for chain and orange for disordered.} \label{Fig:9}
\end{figure*}

As expected, the behaviour of the system from $t=0-100\tau_B$ resembles that of the system in case~3 [see Fig.~\ref{fig:pop_t_preds}(c)] and from $t=100-225\tau_B$ to that of the system in case~1 [see Fig.~\ref{fig:pop_t_preds}(a)]. Interestingly, the change in the aggregation behaviour happens instantaneously: after the parameter change at $t=100\tau_B$, there is an instantaneous drop (increase) in the population fractions of the quadratic, chain and fluid (hexagonal) structures. This rapid structural re-configuration can be attributed to two reasons. The primary reason is that, right before the parameter switch at $t=100\tau_B$, the particles in the system have already formed multiple aggregates of quadratic and chain-like ordering [see Fig.~\ref{Fig:9}(a)]. These aggregates, however, are still separated from each other and there is a lot of empty space around them, enabling rapid reconfiguration. The second reason is that temporal bond autocorrelation functions typically decay faster in hexagonal than in quadratic configurations, as shown in Ref.~\cite{kogler2015generic}. Taken together, after the parameter change, the particles in these aggregates can quickly re-orient themselves to be hexagonally ordered within $1 \tau_B$ [see Figs.~\ref{Fig:9}(b)-(e)]. 

\subsection{Detailed balance condition}   \label{detailed_balance}
\begin{figure*} [htp!]
	\centering
	\includegraphics[width=1.0\textwidth]{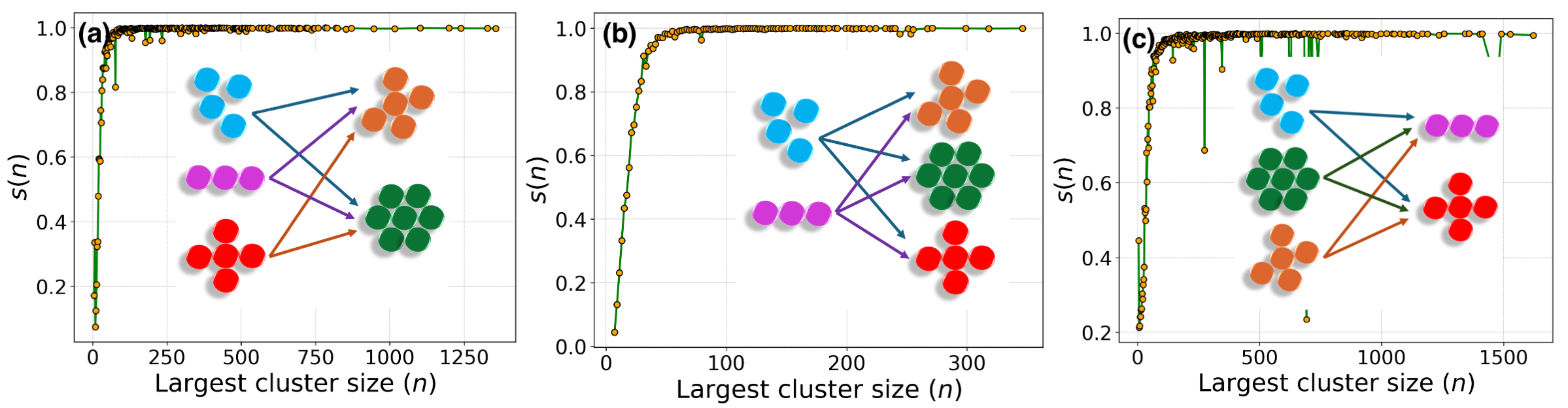}
	\caption{Parameter $s$ as a function of $n$ for (a) case 1, (b) case 2,  and (c) case 3. For each case, the uni-directional transitions are represented by single-sided arrows connecting the relevant states in the insets.} \label{fig:sym_mes}
\end{figure*}

The detailed balance condition states that, in thermal equilibrium, forward and backward transitions between microscopic states are balanced. In classical MSM approaches, the detailed balance condition is typically assumed to hold true and, thus, the transition count matrix $\mathbf{C}$ is explicitly symmetrised (see Section~\ref{construction}). However, the local structural changes during colloidal self-aggregation are non-stationary (Section \ref{handl._stationarity}). Therefore, detailed balance may not necessary hold true throughout the aggregation process. Here, we investigate this issue using our discrete state model. Specifically, we examine the symmetry of the $n-$dependent matrix $\mathbf{C}_n(\tau)$ calculated from 50 different noise realisations. This is done to get a sufficient amount of data for different $n$. Note, however,that $\mathbf{C}_n(\tau)$ is not \textit{averaged} over noise-realisations. To examine the symmetry, we introduce a parameter $s$ by combining the symmetric and anti-symmetric parts of $\mathbf{C}_n(\tau)$ in the following way. 
	
\begin{equation}
	s(n)=\frac{|\mathbf{C}^{s}_{n}(\tau)|-|\mathbf{C}^{as}_{n}(\tau)|}{|\mathbf{C}^{s}_{n}(\tau)|+|\mathbf{C}^{as}_{n}(\tau)|}, \label{eq:sym_mes}
\end{equation}

where $|..|$ represents the Frobenius norm the matrix, $\mathbf{C}^{s}_{n}(\tau)=\frac{1}{2}\left[\mathbf{C}_{n}(\tau)+\mathbf{C}^{\intercal}_{n}(\tau)\right]$ is the symmetric part and $\mathbf{C}^{as}_{n}(\tau)=\frac{1}{2} \left[\mathbf{C}_{n}(\tau)-\mathbf{C}^{\intercal}_{n}(\tau)\right]$ is the anti-symmetric part of $\mathbf{C}_{n}(\tau)$. The parameter $s$ would have values in the range of $s \in [-1, 1]$. A positive value indicates that the matrix is symmetric with the magnitude indicating the extend of symmetricity. A value of $s=+1$ indicates a perfectly symmetric matrix (i.e, $\mathbf{C}^{as}_{n}(\tau)=0$). Likewise, negative values of $s$ indicates that the matrix is anti-symmetric and the magnitude indicates the extent of anti-symmetricity. A value of $s=-1$ indicates a perfectly anti-symmetric matrix. However, since our transition count matrix $\mathbf{C}_n(\tau)$ count the number of transitions, it's elements will not have any negative entries by definition. Therefore none of them will be anti-symmetric. In conclusion, the values for $s$, in our case, would only range between $0\leq s \leq 1$ depending upon the symmetricity. In order to check for the symmetricity of $\mathbf{C}_n(\tau)$ at each $n$, we calculate parameter $s$ for all the three cases. Note that these $s$ values are calculated using $\mathbf{C}_n(\tau)$ before the detailed balance condition is enforced. In Fig.~\ref{fig:sym_mes}, we show the parameter $s$ as a function of $n$ for all the three cases studied here. 

It can be seen that, for all the three cases, the transition count matrices tend to become fully symmetric as $n$ increases. In the initial stages of the aggregation (i.e., $n\leq50$), the matrices are not symmetric, as can be seen from the low values of $s$. In other words, at these early stages, detailed balance is not fulfilled. However, as the aggregation proceeds further ($n>50$), $\mathbf{C}_n(\tau)$ becomes almost perfectly symmetric with $s=+1$ for all the three cases. The violation of detailed balance at early times (i.e., small $n$ values) is caused by uni-directional transitions. These are illustrated in the insets of Fig.~\ref{fig:sym_mes}. One obvious transition that is found in all three cases is the transition from the fluid state (that has been used as the initial configuration) to the other four states. Since our simulations are performed at large coupling strengths ($T^*=0.05$, see Table~\ref{state_points}), particles that enter into one of the four aggregated states (disordered, hexagonal, quadratic or chains) do not transition back into the fluid state. In addition to this, in case 1, all transitions \textit{from} the quadratic and chain states \textit{into} the hexagonal state at low values of $n<50$ are observed to be uni-directional, owing to the preference of hexagonal structures here. At around $n\simeq50$, almost all the particles are in the hexagonal state with no further major changes in population fractions [see Fig.~\ref{fig:CK_sym}(a)]. At even larger $n$ values ($n>100$), we observe small fluctuations in $s(n)$ caused by hexagonal to disordered (and vice-versa) transitions when multiple aggregates merge. However, these transitions obey detailed balance (see Fig.~\ref{fig:sym_mes}(a) at $n\geq100$), thereby maintaining the hexagonal state population fraction close to 1. In case 2, at low value of $n<50$ uni-directional transitions occur \textit{from} the chain state in addition to the fluid state. The particles transition from these two states \textit{into} the other three states [hexagonal, quadratic and disordered, depicted in the inset in Fig.~\ref{fig:sym_mes}(b)] and the remainder of the aggregation process (beyond $n=50$) mainly involves transitions between these three states. From Fig.~\ref{fig:sym_mes}(a), we observe that detailed balance is fulfilled beyond $n=80$ for case 2, indicating that the transitions between the hexagonal, quadratic and disordered states are all equally likely to occur. Finally, for case~3, the transitions originating \textit{from} the hexagonal and disordered states (in addition to the fluid state) in the range $n<50$ are observed to be uni-directional in nature [shown in the inset in Fig.~\ref{fig:sym_mes}(c)]. This is due to the high anisotropy in this case resulting in a strong preference for chain and quadratic structures. Similar to case~2, as aggregation proceeds further ($n>100$), the dynamics are dominated by transitions between the quadratic and chain like structures. From Fig.~\ref{fig:sym_mes}(c), we can see that the detailed balance condition is fulfilled at these $n$ values.

\section{Conclusion}  \label{section: conclusion}
In this study, we have used concepts from Markov state modelling to develop a discrete state model for a self-aggregating colloidal system with anisotropic interactions\cite{kogler2015generic}. The anisotropy originates from the dipole moments induced within the particles by external fields and thus can be tuned. This allows for the formation of aggregates with different structures. Starting from particle-resolved (Brownian dynamics) simulations, we describe the process of self-aggregation as a series of stochastic, memoryless transitions between the different aggregate structures. 

A key characteristic of the current approach, that sets it apart from other MSM approaches for self-aggregation \cite{appeldorn2021employing,husic2018markov,bowman2013introduction,perkett2014using, yang2022nanoparticle, trubiano2022optimization, tang2017construction, sengupta2019automated}, is that it is applicable to systems with hundreds or even thousands of particles which can self-aggregate simultaneously into multiple clusters before merging. In principle, it is possible to construct an MSM where the states also include variables characterizing the global system configurations (despite the challenges of having to deal with multiple, simultaneously evolving aggregates). However, we observed that at the interaction energies and densities considered, the aggregates only merge with one another and a dissolution of entire clusters do not occur. Hence the global configurational transitions of the system are uni-directional in time. Moreover, this process is very slow, taking place over multiple hundreds of $\tau_B$. Therefore, if the states include global configurational information, it can be computationally expensive to obtain sufficient statistics to properly sample the transition probability matrix. Since our primary focus is on the dynamics of local structural transitions, incorporating only local states into the MSM enable us to efficiently capture the dynamics of interest within practical timescales. The global aggregation process is indirectly accounted for via the time-dependence of the largest cluster size, $n$. This approach provides a good balance between predictive capability and computational cost. A recent study \cite{trubiano2024markov} proposed the "MultiMSM" approach for studying such systems where, similar to our approach, the discretization is performed on the local configurational space with time-varying transitions. A difference, however, is that the subunits in our scheme consist of single Brownian particles as opposed to the larger, more complex subunits consisting of multiple particles considered in Ref. \cite{trubiano2024markov}. 

Our study starts from particle-resolved simulation results presented in Ref. \cite{kogler2015generic}. We identify the relevant discrete states based on the orientation of particles in the aggregates. The resulting discretization scheme involves only 5 discrete states (characterized by orientational order parameters), in contrast to the finer discretization schemes used in recent studies \cite{trubiano2024markov, trubiano2022optimization, trubiano2021thermodynamic, perkett2014using}. In agreement with other studies where the discretization is performed on the local configurational space \cite{lander2013crystallization, trubiano2024markov}, the transition rates of our discrete state model are time-dependent in nature. Indeed, we observed significant changes when clusters were merging. After the merging event, the elements of the transition probability matrix were observed to relax to quasi-stationary values. Thus, we found a time-scale separation between the cluster growth (as a result of merging) and transitions between the discrete states (similar to Ref. \cite{lander2013crystallization}). Motivated by this observation, we use, as a key parameter the size of the largest cluster $n$, assuming that the elements of transition probability matrix, $\mathbf{T}_n(\tau)$, relaxes to quasi-stationary values after each change in $n$. The discrete state model is validated by comparing the population fractions calculated from particle-resolved BD simulations with those predicted by the discrete state model as a function of $n$. We then train a Gaussian process regression (GPR) model to infer a continuous function between $\mathbf{T}_n(\tau)$ and $n$. This allows for estimating transition probability matrices for any given value of $n$. With this added flexibility, we were then able to predict population fractions as functions of time, even with noise-average. The present scheme also works in situations where the parameter tuning the anisotropy was changed after a certain time. Finally, we showed, on the coarse-grained scale, that certain uni-directional transitions at the early stages of the aggregation cause a violation of detailed balance. At later stages, detailed balance is maintained, as expected in an equilibrated system.

A limitation of the present approach is that we are restricted to predictions in the range of $n$ values [and the corresponding $\mathbf{T}_n(\tau)$] that has been sampled from BD simulations. Although GPR allows for estimating $\mathbf{T}_n(\tau)$ at any given value of $n$, the estimates become inaccurate if the $n$ value passed as input is out of the range of values that were used to train the GPR model. This is a problem that has been discussed in similar studies \cite{lander2013crystallization, trubiano2021thermodynamic}. The current approach would benefit from techniques that allow to treat such challenges, like adaptive sampling \cite{trubiano2024markov}, and from the use of more comprehensive discretization techniques for identifying the quasi-stationary timescales. This is a subject of future studies. Another interesting research direction would be to develop similar discrete state models for \textit{non-equilibrium} self-aggregating colloidal systems \cite{mallory2019activity, redner2013structure}. For example, recent works have shown that introducing activity to dense systems of triblock Janus colloids can accelerate the aggregation process by helping long-lived transient aggregates to escape kinetic traps \cite{mallory2019activity}. Furthermore, non-reciprocal interactions have also been shown to enhance self-aggregation in certain other colloidal systems \cite{bartnick2016emerging, singh2017non, fehlinger2023collective, navas2024impact}. However, in such systems, the parameter $n$ may not be a slow variable, preventing its use in handling the time-dependent nature of the transition matrix. To handle such systems with the current MSM framework, other slow evolving order parameters need to be identified.
 
The MSM approach described in this work, in its present version, is data-driven and reliant on BD simulations. It would indeed be interesting, from a computational efficiency perspective, to develop coarse-graining strategies that can function independently of particle-resolved simulations as a stand-alone technique. Extending our approach using techniques such as adaptive sampling could potentially be used to predict dynamics and long-time behaviour without the added computational complexity of performing BD simulations for the entire temporal range. In adaptive sampling techniques, multiple short BD simulations started from different initial configurations are used to build the overall MSM \cite{trubiano2024markov}. Additionally, our MSM approach could potentially also be extended to predict population fraction dynamics at different number densities, without performing any additional BD simulations. A similar prediction method was proposed Ref.~ \cite{trubiano2024markov}.

Although our MSM model is focussing on local structural transitions, we still account for changes in the global structural transitions by monitoring the parameter $n$. In this sense, we implicitly retain some information about the global spatial configuration in our MSM. The algorithm that we use to determine the value of $n$, enables us to have information regarding the cluster size distribution of the entire system. So in principle, conversion back to a particle-based view may be obtained with information regarding the population fraction in each of the states and also information regarding the cluster distribution. This information can be used to \textit{generate} the corresponding particle-resolved system configurations. However, accuracy regarding the distances between the different aggregates and their orientation cannot be guaranteed, since such information is not available on the coarse-grained MSM level. Further research in these directions could pave the way for broader applicability. Finally, from a conceptual point of view, it might be interesting to compare non-equilibrium thermodynamic properties on the coarse-grained versus the particle-resolved scale.

\begin{acknowledgements}
	This work was funded by the Deutsche Forschungsgemeinschaft (DFG, German Research Foundation), project number 449485571. 
\end{acknowledgements}

\section*{Author declarations}
\subsection*{Conflict of Interest}
The authors have no conflicts to disclose. 

\subsection*{Author contributions}
\textbf{Salman Fariz Navas:} Formal analysis (lead); software (lead); writing - original draft preparation (lead). \textbf{Sabine H. L. Klapp}: Conceptualization (lead); writing – review and editing (lead), funding acquisition (lead).

\section*{Data Availability Statement}
The data that support the findings of this study are available from the corresponding authors upon reasonable request.

\appendix
\section{Normalization of $U_{\text{DIP}}(\mathbf{r}_{ij})$} \label{app_norm_udip}
The interaction potential $U_{\text{DIP}}(\mathbf{r}_{ij})$ has to be normalised such that it's value remains the same at the minima, irrespective of the value of $\delta_r$. The minima of the potential occurs when two particles are at contact ($r_{ij}=\sigma$) and the vector $\mathbf{r}_{ij}$ is pointing along one of the fields ($\mathbf{r}_{ij}=\sigma\mathbf{\hat{e}}_{\alpha}$). The normalization is done in the following way,

\begin{equation}
	\tilde{U}_{\text{DIP}}(\mathbf{r}_{ij})=U_{\text{DIP}}(\mathbf{r}_{ij})\frac{u}{U_{\text{DIP}}(\sigma \mathbf{\hat{e}}_{\alpha})}, \label{eq:U_DIP_norm}
\end{equation}
	
where $u=-0.2804\epsilon$  is a constant (chosen consistently with Ref. \cite{kogler2015generic}) calculated from the un-normalized potential $U_{\text{DIP}}(\sigma \mathbf{\hat{e}}_{\alpha})$ for $\delta_r=0.3\sigma$.

\section{Gaussian process regression} \label{appendix: gpr}
Gaussian process regression (GPR) is a non-linear, non-parametric regression technique used for interpolating existing data \cite{williams2006gaussian}. It is particularly useful when the analytic function describing the underlying process is unknown and the observed data are very noisy. GPR is different from other parametric regression techniques using closed functional forms like linear regression or cubic splines, in the sense that it does not assume any \textit{ansatz}. It solely relies on large amounts of data (observations) to estimate a function, from which missing data are predicted. This is done by assuming a distribution of functions described by a Gaussian process, that is, a generalization of the Gaussian distribution to function spaces. While Gaussian distributions describe the properties of random variables (scalars or vectors), Gaussian processes describe the properties of functions.

Consider that we have a dataset $\{(x_i,y_i)\}_{i=1}^{n}$ where $x_i$ are the input points and $y_i$ are the observations. We intend to estimate the function $f(x)$ mapping $x_i$ to $y_i$. It is also assumed that the observations ($y_i$) are subject to noise as well, giving $y_i=f(x_i)+\epsilon_i$. Here, $\epsilon_i$ represents a Gaussian distribution with zero mean and $\sigma_n^2$ variance. To estimate the underlying function $f(x)$, we start by assuming a distribution of functions (a Gaussian process) from which $f(x)$ can potentially be sampled. Gaussian processes are described by a mean function $m(x)=\E[f(x)] $ and a covariance function $k(x,x')=\E[(f(x)-m(x))(f(x')-m(x'))] $ (commonly referred to as the kernel). These properties are the same for all $f(x)$ that are generated by a particular Gaussian process. For GPR, $m(x)$ is usually assumed to be zero, by default, and a suitable kernel $k(x,x')$ can be chosen. We use GPR in our study to estimate the function mapping $\mathbf{T}_n(\tau)$ to $n$. Although we expect the elements of $\mathbf{T}_n(\tau)$ to vary in a discontinuous manner as a function of $n$, we expect the variance to be rather small across nearby values of $n$. A suitable kernel that can handle such data is the Ornstein-Uhlenbeck kernel (i.e. the covariance function of the Ornstein-Uhlenbeck process \cite{uhlenbeck1930theory, williams2006gaussian}) which is given by
\begin{equation}
	k_{\text{O.U}}(n,n')=\exp\left(-\frac{|n - n'|}{l}\right),
\end{equation}

where the parameter $l$ corresponds to the correlation length. Small values of $l$ mean that the kernel is particularly strong at smaller distances of $n$ (i.e., $f(x)$ can vary more rapidly with change in $n$), while larger values of $l$ mean that the covariance is stronger for larger distances of $n$ (i.e., $f(x)$ varies more gradually with change in $n$). The kernel is used to define a prior probability distribution of functions encapsulating our assumptions about the underlying true function, called the Gaussian process prior. Using Bayes' theorem, this prior distribution along with our observations (here: the $\mathbf{T}_n(\tau)$ sampled at different $n$ from particle resolved simulations) is used to estimate a narrower posterior distribution of functions which are likely to fit our observations. This is done by optimizing the parameters of the kernel, namely, $l$. The parameters are optimized separately for each of the 25 matrix elements. We can then obtain the predictive mean from the posterior distribution which gives the expected value of the \textit{estimated} function at new input points. In the present calculations, we have found that the predictions for the $\delta_r=0.21\sigma$ case were significantly improved when a composite kernel combining $k_{\text{O.U}}(n,n')$ and a white noise kernel $k_{\text{white}}(n,n')=\sigma_n^2\delta(n,n')$ were combined,
\begin{equation}
	k(n,n')=\exp\left(-\frac{|n - n'|}{l}\right)+\sigma_n^2\delta(n,n'),
\end{equation}

where the parameter $\sigma_n$ is the variance of the white noise. Hence, for $\delta_r=0.21\sigma$, the parameters $l$, as well as $\sigma_n$ are optimized. We use the GPR implementation in the python package \textit{scikit-learn} \cite{scikit-learn} for our analyses. The values of the optimized kernel parameters (parameter $l$ for cases 1,2 and 3 and parameter $\sigma_n$ for case 2 only) for the three cases are given below, for each of the matrix elements. 

\subsection{Case 1}
\[
\mathbf{L} = \resizebox{0.9\columnwidth}{!}{$
	\begin{bmatrix}
	1.03\times 10^{3} & 9.48\times10^{03} & 1.26\times10^{6} & 3.54\times10^{5} & 1.10\times10^{7} \\
	5.67\times10^{5} & 9.654\times10^{2} & 8.61\times10^{6} & 6.30\times10^{6} & 3.72\times10^{7} \\
	9.06\times10^{4} & 5.58\times10^{4} & 1.50\times10^3 & 1.89\times10^5 & 2.43\times10^4 \\
	5.43\times10^{5} & 1.42\times10^{5} & 8.90\times10^{4} & 1.25\times10^{3} & 3.84\times10^{7} \\
	6.50\times10^{5} & 1.30\times10^{4} & 3.44\times10^{4} & 7.45\times10^{5} & 1.41\times10^{3}
\end{bmatrix}
$}
\]

\subsection{Case 2}
\[
\mathbf{L} = \resizebox{0.9\columnwidth}{!}{$
	\begin{bmatrix}
	2.32\times10^{4} & 8.62\times10^{6} & 1.37\times10^{8} & 6.33\times10^{9} &8.261\times10^8\\ 
	1.91\times10^6 & 2.53\times10^4 & 5.02\times10^6 & 1.68\times10^7 & 1.10\times10^8 \\
	1.64\times10^6 & 5.44\times10^6 & 2.30\times10^4 & 4.12\times10^7 & 4.13\times10^6 \\
	4.829\times10^6 & 1.46\times10^6 & 3.73\times10^6 & 2.90\times10^4 & 1.38\times10^8 \\
	1.20\times10^6 & 1.18\times10^6 & 2.41\times10^6 & 2.05\times10^8 & 1.56\times10^4
\end{bmatrix}
$}
\]

\[
\mathbf{\sigma_n} = \resizebox{0.9\columnwidth}{!}{$
	\begin{bmatrix}
	1.44\times10^{-3}& 3.20\times10^{-5}& 4.01\times10^{-8}& 5.51\times10^{-9}& 6.23\times10^{-9} \\
	3.96\times10^{-6}& 1.47\times10^{-3}& 9.06\times10^{-7}& 8.94\times10^{-7}& 1.01\times10^{-7} \\
	3.34\times10^{-6}& 3.15\times10^{-6}& 1.43\times10^{-3}& 9.60\times10^{-7}& 6.89\times10^{-7} \\
	6.47\times10^{-7}& 1.23\times10^{-5}& 6.72\times10^{-6}& 1.27\times10^{-3}& 9.82\times10^{-8}\\
	3.18\times10^{-6}& 1.19\times10^{-4}& 4.65\times10^{-5}& 2.41\times10^{-7}& 8.86\times10^{-4}
\end{bmatrix}
$}
\]

\subsection{Case 3}
\[
\mathbf{L} = \resizebox{0.9\columnwidth}{!}{$
	\begin{bmatrix}
		1.83\times10^3 & 9.76\times10^7 & 1.06\times10^8 & 1.00\times10^10 & 7.39\times10^8 \\
	1.03\times10^5 & 3.37\times10^3 & 4.06\times10^4 & 8.46\times10^6 & 4.87\times10^5 \\
	8.94\times10^5 & 4.69\times10^6 & 1.89\times10^3 & 3.09\times10^7 & 8.68\times10^6 \\
	1.36\times10^6 & 5.72\times10^5 & 2.87\times10^4 & 3.01\times10^3 & 2.07\times10^6 \\
	6.02\times10^5 & 5.54\times10^6 & 1.04\times10^6 & 7.82\times10^7 & 1.91\times10^3
\end{bmatrix}
$}
\]

\section{Transition probability matrix as a function of $n$}
\begin{figure*} [htp]
	\centering
	\includegraphics[width=1.0\textwidth]{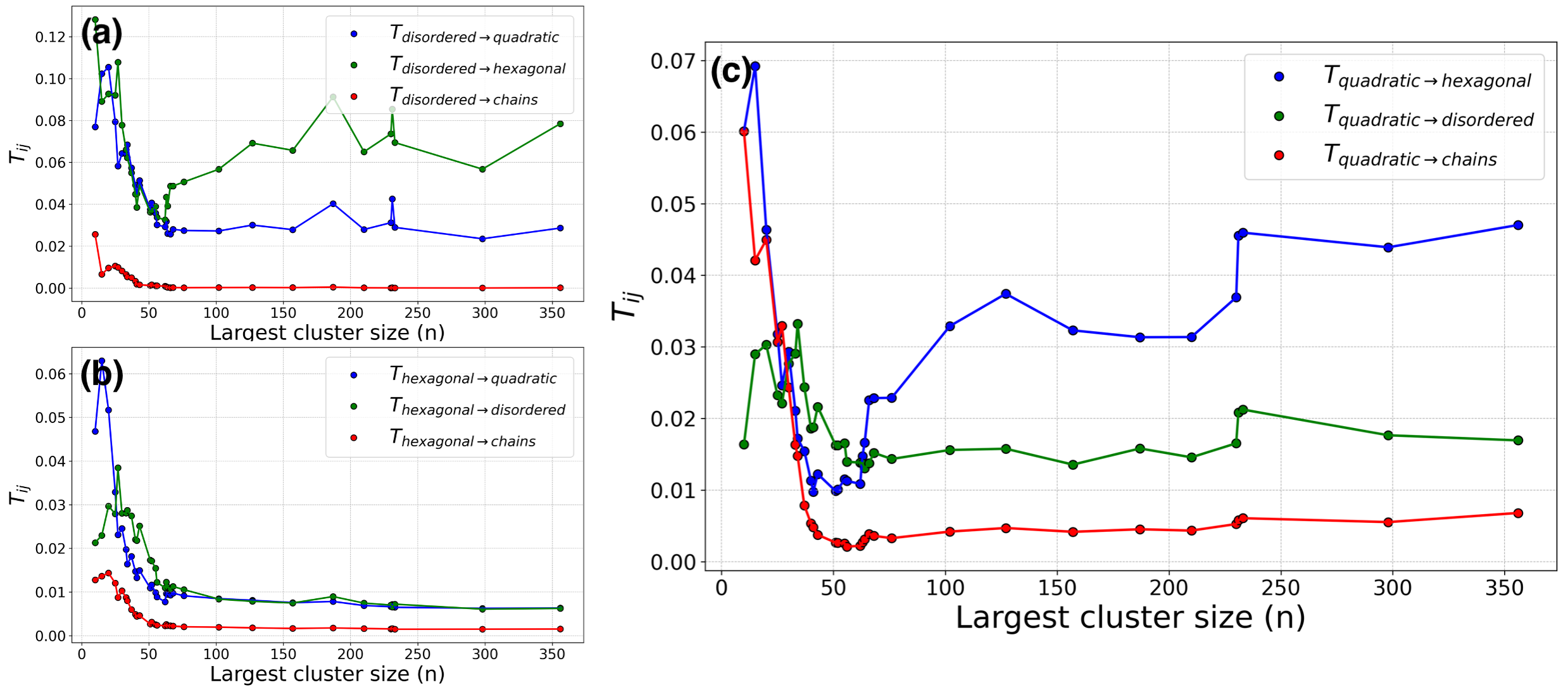}
	\caption{Elements of the transition probability matrix corresponding to transitions originating from (a) disordered (b) hexagonal and (c) quadratic states as a function of $n$ for a single noise realization of the system studied in case~2} \label{fig:appendix}
\end{figure*}

In Fig.~\ref{fig:appendix}, we show, as an example, elements of the transition probability matrix corresponding to transitions between the disordered, hexagonal, quadratic and chain states, for the system in case~2, for a single noise realization.

\bibliography{references.bib}

\end{document}